\documentclass[showpacs,aps,prd,reprint,superscriptaddress,nofootinbib,longbibliography]{revtex4-1}

\usepackage[colorlinks=true, pdfstartview=FitV, bookmarks=true, bookmarksnumbered=true, breaklinks]{hyperref}
\usepackage[dvipdfmx]{graphicx}
\usepackage{amsmath,amssymb,bm,color,longtable,mathrsfs,amsfonts}

\definecolor{DeepPink2}{rgb}{0.932,0.07,0.536}
\definecolor{RoyalBlue1}{rgb}{0.284,0.464,1}
\definecolor{SpringGreen3}{rgb}{0,0.804,0.4}
\hypersetup{linkcolor=RoyalBlue1, citecolor=DeepPink2, urlcolor=SpringGreen3}

\begin{document}
\begin{flushright}
\end{flushright}

\title{Doubly heavy tetraquarks in a chiral-diquark picture}

\author{Yonghee~Kim}
\email[]{\it kimu.ryonhi@phys.kyushu-u.ac.jp}
\affiliation{Department of Physics, Kyushu University, Fukuoka 819-0395, Japan}


\author{Makoto~Oka}
\email[]{\it oka@post.j-parc.jp}
\affiliation{Advanced Science Research Center, Japan Atomic Energy Agency (JAEA), Tokai 319-1195, Japan}
\affiliation{Nishina Center for Accelerator-Based Science, RIKEN, Wako 351-0198, Japan}

\author{Kei~Suzuki}
\email[]{{\it k.suzuki.2010@th.phys.titech.ac.jp}}
\affiliation{Advanced Science Research Center, Japan Atomic Energy Agency (JAEA), Tokai 319-1195, Japan}

\date{\today}

\begin{abstract}
Energy spectrum of doubly heavy tetraquarks, $T_{QQ}$ ($QQ \bar{q} \bar{q}$ with $Q = c, b$ and $q = u, d, s$), is studied in the potential chiral-diquark model. Using the chiral effective theory of diquarks and the quark-diquark-based potential model, the  $T_{bb}$, $T_{cc}$, and $T_{cb}$ tetraquarks are described as a three-body system composed of two heavy quarks and an antidiquark. We find several $T_{bb}$ bound states, while no $T_{cc}$ and $T_{cb}$ (deep) bound state is seen. We also study the change of the $T_{QQ}$ tetraquark masses under restoration of chiral symmetry. 
\end{abstract}
\maketitle

\section{Introduction}\label{Sec_1}

Hadron spectroscopy is one of the basis to show the properties of the quantum field theory.  Hadrons are mostly observed as two types of structures, the quark-antiquark configuration for the conventional mesons and the three-quark configuration for the conventional baryons. However, the Quantum Chromodinamics (QCD) allows some exotic color-singlet hadrons such as compact multiquark states, hadronic molecules, and glueballs.
Observation and interpretation of these unconventional states with the special properties have provided us a various attempt for the discussion of hadron physics for a long time. 

A recent discovery of exotic hadrons is the $T_{cc}^+$ tetraquark reported by the LHCb Collaboration \cite{LHCb:2021vvq, LHCb:2021auc}. This is composed of two charm quarks and two light antiquarks, and the total spin and parity is assumed to be $J^P=1^+$.
The mass relative to $D^{*+}-D^0$ threshold and the decay width were firstly determined as $-273\pm 61$ keV and $410\pm 165$ keV with the Breit-Wigner profile \cite{LHCb:2021vvq}, and these values have been updated as $-361\pm 40$ keV and $47.8\pm 1.9$ keV with the unitarized Breit-Wigner profile \cite{LHCb:2021auc}, respectively. 
On the theoretical points of view, the early studies of exotic hadrons including two heavy quarks, $QQ\bar q \bar q$ ($Q=c, b$ and $q=u, d, s$), have been made in the 1980s \cite{Ader:1981db, Ballot:1983iv, Lipkin:1986dw, Zouzou:1986qh, Heller:1986bt, Carlson:1987hh}. Until these days, many researchers attempted to search for the features of doubly heavy $T_{QQ}$ tetraquarks with various approaches such as 
quark-level models \cite{Silvestre-Brac:1993zem, Semay:1994ht, Chow:1994mu, Chow:1994hg, Pepin:1996id, Brink:1998as, Gelman:2002wf, Vijande:2003ki, Janc:2004qn, Vijande:2006jf, Ebert:2007rn, Zhang:2007mu, Vijande:2009kj, Yang:2009zzp, Feng:2013kea, Karliner:2017qjm, tetraeichten, Yan:2018gik, Ali:2018xfq, Park:2018wjk, Deng:2018kly, Carames:2018tpe, Hernandez:2019eox, Yang:2019itm, Bedolla:2019zwg, Yu:2019sxx, Wallbott:2020jzh, Tan:2020ldi, Lu:2020rog, tetrabraaten, Yang:2020fou, 
Qin:2020zlg, 
tetrahiyama, Meng:2021yjr, Chen:2021tnn, Jin:2021cxj, Chen:2021cfl, Andreev:2021eyj, Deng:2021gnb}, 
chromomagnetic interaction models \cite{Lee:2007tn, Lee:2009rt, Hyodo:2012pm, Luo:2017eub, Hyodo:2017hue, Cheng:2020nho, Cheng:2020wxa, Weng:2021hje, Guo:2021yws}, 
QCD sum rules \cite{Navarra:2007yw, Dias:2011mi, Du:2012wp, Chen:2013aba, Wang:2017uld, Wang:2017dtg, Agaev:2018khe, Sundu:2019feu, Agaev:2019kkz, Tang:2019nwv, Agaev:2019lwh, Agaev:2020dba, Wang:2020jgb, Agaev:2020zag, Agaev:2020mqq, Agaev:2021vur, Azizi:2021aib, Aliev:2021dgx, Ozdem:2021hmk, Bilmis:2021rdp}, 
lattice QCD simulations \cite{Detmold:2007wk, Wagner:2010ad, Bali:2010xa, Bicudo:2012qt, Brown:2012tm, Ikeda:2013vwa, Bicudo:2015vta, Bicudo:2015kna,  Francis:2016hui, Bicudo:2016ooe, Bicudo:2017szl, Francis:2018jyb, tetralatticebbcc, Leskovec:2019ioa, tetralatticebc, Mohanta:2020eed, Bicudo:2021qxj, Padmanath:2021qje} and so on. 
Most of them predict a stable doubly-bottom $T_{bb}$ tetraquark below the strong-decay threshold, but no stable doubly charmed $T_{cc}$ tetraquark. The predictions for the bottom-charmed $T_{cb}$ tetraquark are divided, whether it is stable or not.

Multiquark states are often considered from the perspective of their substructures (clusters). The $T_{QQ}$ tetraquarks are often discussed as diquark-antidiquark or meson-meson structure. In this paper, we apply the chiral effective theory of diquarks and the quark-diquark potential model \cite{CETdiquark, scalar, Vdiquark}. 
In Ref.~\cite{CETdiquark}, we proposed a chiral effective theory of the spin 0,  color $\bf{\bar 3}$, scalar/pseudoscalar diquarks based on the $SU(3)_R\times SU(3)_L$ chiral symmetry. Here, the $0^{\pm}$ diquarks belong to the same chiral multiplet, forming chiral partners and their masses are to be degenerate when the chiral symmetry is restored. 
Similarly, in Ref.~\cite{Vdiquark}, the spin 1, color $\bf{\bar 3}$, vector/axial-vector diquarks are introduced as another set of chiral partners. Then we applied the chiral effective theories to singly heavy baryons and studied the spectrum using the nonrelativistic potential heavy-quark--diquark model.

In the present work, $T_{QQ}$ tetraquarks are described as a three-body system composed of two heavy quarks and an antidiquark. We calculate the mass spectrum and the wave functions to study how chiral symmetry is realized in the tetraquark systems. We also investigate the dependence of $T_{QQ}$ tetraquark masses on the chiral symmetry breaking parameter. 

This paper is organized as follows.
In Sec.~\ref{Sec_2}, we introduce the chiral effective theory of diquarks and construct the potential models for the diquark-pictured $T_{QQ}$ tetraquarks.
In Sec.~\ref{Sec_3}, we show the results for various $T_{QQ}$ tetraquark states.
The last section \ref{Sec_4} is for our conclusion and outlook.  

\section{Theoretical Framework} \label{Sec_2}

In this section, firstly, we introduce the mass formulas of diquarks given by the chiral effective theory of scalar/pseudoscalar diquarks \cite{CETdiquark, scalar} and of vector/axial-vector diquarks \cite{Vdiquark}. Then we construct the nonrelativistic potential model for the $T_{QQ}$ tetraquark states including the antidiquark cluster.
\subsection{Chiral effective theory of diquarks}
  

Following Ref. \cite{Vdiquark}, we consider the chiral effective theory of diquarks given by the chiral effective Lagrangian,
\begin{eqnarray}
&&\mathcal{L}=\mathcal{L}_S + \mathcal{L}_V + \frac{1}{4}{\rm Tr}[\partial^\mu \Sigma^\dag \partial_\mu \Sigma]-V(\Sigma)+\mathcal{L}_{V-S}. 
\label{totlag}
\end{eqnarray}
Here the first term $\mathcal{L}_S$ describes the effective Lagrangian for the scalar and pseudoscalar diquarks, and $\mathcal{L}_V$ for the vector and axial-vector diquarks. 
Their explicit forms are given below. The last term, $\mathcal{L}_{V-S}$, describes the coupling between the scalar and vector diquarks. As it does not contribute to the diquark masses, here we omit this term.
The third and forth terms of Eq.~(\ref{totlag}) are the kinetic and potential terms for the meson field $\Sigma$, respectively. This meson operator $\Sigma$ contains nonet scalar $\sigma$ and pseudoscalar $\pi$ mesons, whose chiral transform is given by 
\begin{equation}
\Sigma_{ij}=\sigma_{ij}+i\pi_{ij} \rightarrow U_{L,ik} \Sigma_{km} U^{\dag}_{R,mj}.
\end{equation}
The potential term for the $\Sigma$ meson field, $V(\Sigma)$, causes spontaneous chiral symmetry breaking, represented by the vacuum expectation value of scalar meson $\sigma$ as
\begin{equation}
\langle \Sigma_{ij} \rangle =\langle \sigma_{ij} \rangle =f_{\pi} \delta_{ij}~,~~~\langle \pi_{ij} \rangle =0,
\label{vevmeson}
\end{equation}
where $f_\pi \simeq 92$ MeV is the pion decay constant. In this vacuum, $\pi$ meson is regarded as the massless Nambu-Goldstone bosons. 

Using the $\Sigma$ meson field and the chiral diquark operators summarized in Table.~\ref{diopchiral}, the effective Lagrangians, $\mathcal{L}_S$ and $\mathcal{L}_V$, are expressed as
\begin{eqnarray}
\begin{split}
\mathcal{L}_S&=\mathcal{D}_\mu d_{R,i}(\mathcal{D}^\mu d_{R,i})^\dag + \mathcal{D}_\mu d_{L,i}(\mathcal{D}^\mu d_{L,i})^\dag \\
		   &-m_{S0}^2(d_{R,i}d_{R,i}^\dag+d_{L,i}d_{L,i}^\dag) \\
		   &-\frac{m_{S1}^2}{f_\pi}(d_{R,i}\Sigma_{ij}^\dag d_{L,j}^\dag+d_{L,i}\Sigma_{ij}d_{R,j}^\dag) \\
		   &-\frac{m_{S2}^2}{2f_\pi^2}\epsilon_{ijk}\epsilon_{lmn} (d_{R,k}\Sigma_{li}\Sigma_{mj}d_{L,n}^\dag+d_{L,k}\Sigma^\dag_{li}\Sigma^\dag_{m,j}d_{R,n}^\dag), \\
\end{split}
\label{slag}
\end{eqnarray}
\begin{eqnarray}
\begin{split}
\mathcal{L}_V&=\frac{1}{2}{\rm Tr}[F^{\mu\nu}F^{\dag}_{\mu\nu}]
                    +m_{V0}^2{\rm Tr}[d^{\mu}d^{\dag}_{\mu}] \\
                    &+\frac{m_{V1}^2}{f_{\pi}^{2}}{\rm Tr}[\Sigma^{\dag}d^{\mu}\Sigma^Td^{\dag T}_{\mu}] \\
                    &+\frac{m_{V2}^2}{f_{\pi}^{2}}[{\rm Tr}\{ \Sigma^{T}\Sigma^{\dag T} d_{\mu}^{\dag}d^{\mu} \}
                    					     +{\rm Tr}\{ \Sigma\Sigma^{\dag} d^{\mu}d^{\dag}_{\mu} \}]. \\
\end{split}
\label{vlag}
\end{eqnarray}

 \begin{table}[htb]
  \centering
    \caption{Diquark operators in the chiral basis.}   
  \begin{tabular}{ l | c | c | c  } \hline\hline
   Chiral operator & Spin & Color & Flavor \\ \hline \hline
 $d^a_{R,i} = \epsilon_{abc}\epsilon_{ijk}(q^{bT}_{R,j} Cq^c_{R,k})$ &
 										$0$&$\overline{\bm 3}$&$\bar{\bm 3}$\\ 
 $d^a_{L,i} = \epsilon_{abc}\epsilon_{ijk}(q^{bT}_{L,j} Cq^c_{L,k})$ &
                                                                                    $0$&$\overline{\bm 3}$&$\bar{\bm 3}$\\ 
 $d^{a,\mu}_{ij} = \epsilon_{abc}(q^{bT}_{L,i} C \gamma^\mu q^c_{R,j})$ &
                                                                                         $1$&$\overline{\bm 3}$&$\bar{\bm 3}, \bm{6}$\\  \hline \hline
  \end{tabular}  
  \label{diopchiral}
  \end{table}

Since the diquark is not a color-singlet state, we introduce the color-gauge-covariant derivative, $\mathcal{D}^{\mu}=\partial^{\mu}+igT^{\alpha}G^{\alpha,\mu}$, with $G^{\alpha,\mu}$ being the gluon field and $T^\alpha$ being the color $SU(3)$ generator for the color $\bf \bar{3}$ representation. It is used for the kinetic term of these Lagrangians. $F^{\mu \nu}=\mathcal{D}^{\mu}d^{\nu}-\mathcal{D}^{\nu}d^{\mu}$ in the first term of Eq.~(\ref{vlag}) shows the strength of chiral vector diquark fields. The self-interaction terms of gluons are omitted, and all the color indices are contracted and not explicitly written. 

The masses of scalar and pseudoscalar diquarks are given by the three parameters $m_{S0}^2$, $m_{S1}^2$, and $m_{S2}^2$ in Eq.~(\ref{slag}). $m_{S0}$ is the chiral invariant mass which is the degenerate mass for the two diquarks in the chiral symmetry restored phase. Mass splitting between diquarks induced by the chiral symmetry breaking is given by $m_{S1}$ and $m_{S2}$. In particular, the $m_{S1}^2$ term also causes the $U_A(1)$ symmetry breaking \cite{tHooft:1976snw, tHooft:1986ooh}. On the other hand, the masses of vector and axial-vector diquarks are given by three parameters $m_{V0}^2$, $m_{V1}^2$, and $m_{V2}^2$ in Eq.~(\ref{vlag}). Similar to the scalar and pseudoscalar diquarks, $m_{V0}$ is the chiral invariant mass of vector and axial-vector diquarks, and their mass splitting is given by $m_{V1}$ and $m_{V2}$.

The non-zero strange quark mass brings about the explicit chiral symmetry breaking and the flavor $SU(3)$ symmetry breaking. In this term, the masses of scalar and pseudoscalar diquarks, $M(0^+)$ and $M(0^-)$, are expressed in Ref.~\cite{CETdiquark} as
\begin{eqnarray}
\left[M_{ud} (0^+)\right]^2&&= m_{S0}^2 - (1+\epsilon) m_{S1}^2 - m_{S2}^2, \label{Smass1} \\
 \left[M_{ns} (0^+)\right]^2&&= m_{S0}^2 - m_{S1}^2 - (1+\epsilon) m_{S2}^2, \label{Smass2} \\
 \left[M_{ud} (0^-)\right]^2&&= m_{S0}^2 + (1+\epsilon) m_{S1}^2 + m_{S2}^2, \label{Smass3} \\
 \left[M_{ns} (0^-)\right]^2&&= m_{S0}^2 + m_{S1}^2 + (1+\epsilon) m_{S2}^2, \label{Smass4} 
\end{eqnarray}
where the index $n$ stands for $u$ or $d$ quark, and $s$ for $s$ quark, meaning two constituent quarks of a diquark.
The constant $\epsilon$ is given by
\begin{eqnarray}
\epsilon = \frac{f_s}{f_\pi} \left( 1+\frac{m_s-g_sf_\pi}{g_s f_s} \right) \simeq 2/3,
\end{eqnarray}
with $f_s=2f_K-f_\pi$, where $m_s, g_s, f_\pi$, and $f_K$ stands for the mass of strange quark, the quark-meson coupling constant, and the decay constants of pion and kaon, respectively.
From these mass formulas, Eqs.~(\ref{Smass1})--(\ref{Smass4}), we can give the mass relation of scalar and pseudoscalar diquarks \cite{CETdiquark, scalar, Vdiquark} as
\begin{eqnarray}
[M_{ns}(0^+)]^2 - [M_{ud}(0^+)]^2 = [M_{ud}(0^-)]^2 - &&[M_{ns}(0^-)]^2 \nonumber \\
												&&> 0.
\label{inverse}
\end{eqnarray}
From Eq.~(\ref{inverse}), we see the non-strange pseudoscalar diquark is heavier than the singly strange pseudoscalar diquark, $M_{ud}(0^-) > M_{ns}(0^-)$, which is called the inverse mass hierarchy. 

The masses of axial-vector and vector diquarks, $M(1^+)$ and $M(1^-)$, are expressed in Ref.~\cite{Vdiquark} as
\begin{eqnarray}
 \left[M_{nn} (1^+)\right]^2&&= m_{V0}^2 + m_{V1}^2 + 2m_{V2}^2, \label{Vmass1} \\
 \left[M_{ns} (1^+)\right]^2&&= m_{V0}^2 + m_{V1}^2 + 2m_{V2}^2 \nonumber \\
 &&+\epsilon(m_{V1}^2+2m_{V2}^2), \label{Vmass2} \\
 \left[M_{ss} (1^+)\right]^2&&= m_{V0}^2 + m_{V1}^2 + 2m_{V2}^2 \nonumber \\
  &&+2\epsilon(m_{V1}^2+2m_{V2}^2), \label{Vmass3} \\
 \left[M_{ud} (1^-)\right]^2&&= m_{V0}^2 - m_{V1}^2 + 2m_{V2}^2, \label{Vmass4} \\
 \left[M_{ns} (1^-)\right]^2&&= m_{V0}^2 -m_{V1}^2 + 2m_{V2}^2 \nonumber \\
  &&+\epsilon(-m_{V1}^2+2m_{V2}^2). \label{Vmass5}
\end{eqnarray} 
For the masses of axial-vector diquarks, Eqs.~(\ref{Vmass1})--(\ref{Vmass3}), the generalized Gell-Mann--Okubo mass formula \cite{Vdiquark} which is analogous to the conventional one \cite{gomass1, gomass2} is obtained as
\begin{eqnarray}
 [M_{ss} (1^+)]^2 -  [M_{ns} (1^+)]^2 &&=  [M_{ns} (1^+)]^2 -  [M_{nn} (1^+)]^2 \nonumber\\
&&= \epsilon (m_{V1}^2+2m_{V2}^2). 
\label{axialgomass}
\end{eqnarray} 
Here the square mass differences are characterized by the number of $s$ quark. 
On the other hand, the square mass difference between nonstrange and singly strange vector diquarks is given by \begin{eqnarray}
 [M_{ns} (1^-)]^2 -  [M_{ud} (1^-)]^2 = \epsilon(-m_{V1}^2+2m_{V2}^2). 
\label{vec-sqmassdiff}
\end{eqnarray}
As shown in Table. \ref{diquarkmass}, the parameter $m_{V1}^2$ generally takes the negative value \cite{Vdiquark}, so that the mass difference between nonstrange and singly strange vector diquarks becomes much larger than that of axial-vector diquarks. Here we call this the enhanced mass hierarchy of vector diquarks in this paper.

\begin{table}[tb]
  \centering
    \caption{Diquark masses and parameters of the chiral effective Lagrangian taken from Refs.~\cite{scalar, Vdiquark}, whose values are determined by the Y-potential model. } 
  \begin{tabular}{ l   c  | c  r} \hline\hline 
\multicolumn{2}{ l |}{ Diquark masses ($\rm MeV$)} 
&\multicolumn{2}{ l }{Parameters  ({$\rm MeV^2$})} \\ \hline
    $ M_{ud} (0^+)~ $ & 725  &$~~m_{S0}^2~$ &$ (1119)^2$  \\
    $ M_{ns} (0^+)~ $ & 942  &$~~m_{S1}^2~$ &$ (690)^2$  \\
    $ M_{ud} (0^-)~ $ & 1406 &$~~m_{S2}^2~$ &$-(258)^2$    \\ 
    $ M_{ns} (0^-)~ $ & 1271  &&  \\  \hline   
    $ M_{nn} (1^+)~ $ & 973    &$~~m_{V0}^2~$ &$ (708)^2$ \\
    $ M_{ns} (1^+)~ $ & 1116   &$~~m_{V1}^2~$ &$-(757)^2$ \\
    $ M_{ss} (1^+)~ $ & 1242   &$~~m_{V2}^2~$ &$ (714)^2$ \\ 
    $ M_{ud} (1^-)~ $ & 1447   && \\ 
    $ M_{ns} (1^-)~ $ & 1776  &&\\  \hline\hline     
  \end{tabular}
  \label{diquarkmass}
\end{table}

\subsection{Potential models}

 In order to calculate the spectrum of $T_{QQ}$ tetraquarks, we use the chiral effective theory of diquark \cite{CETdiquark, scalar, Vdiquark} and consider these tetraquarks as the three-body system with two heavy quarks ($Q_1, Q_2$) and one antidiquark ($\bar d$). 
We need two kinds of two-body potentials, one for the interaction between two heavy quarks, $V_{Q_1 Q_2}$, and the other for between a heavy quark and a color $\bm 3$ antidiquark, $V_{Q_i \bar d}$. 
 
\begin{figure}[htbp]
  \includegraphics[clip,width=1.0\columnwidth]{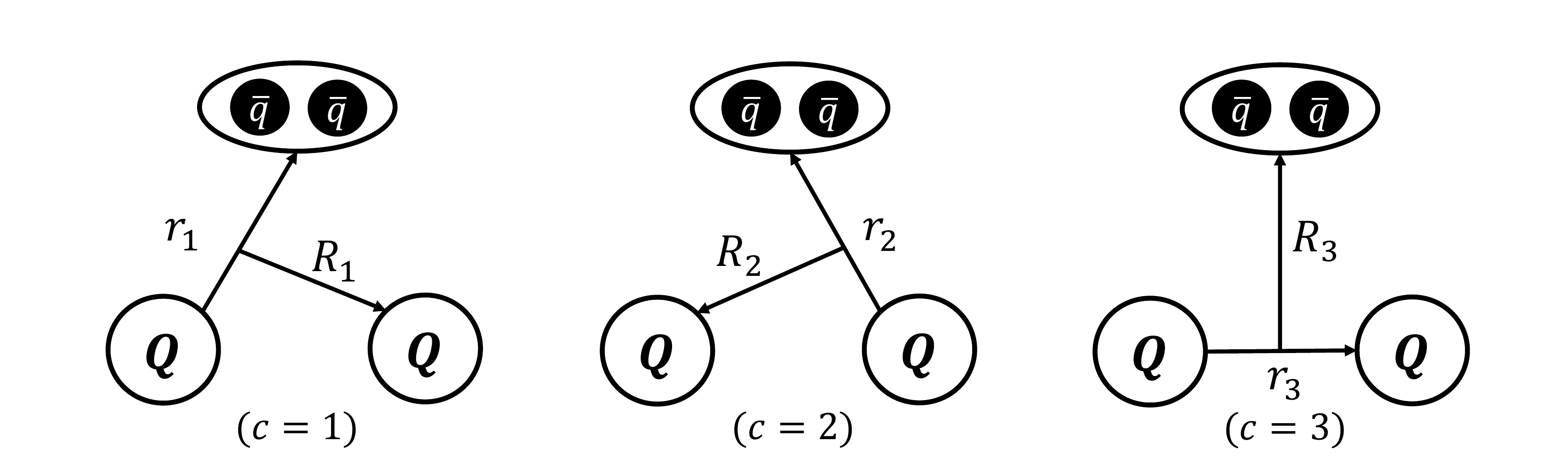}
  \caption{Jacobi coordinates for the three-body system. In each channel ($c=1, 2, 3$), $Q$ and $\bar q$ denote the heavy quark and the light antiquark, respectively.}
  \label{jacobi}
\end{figure}

The Hamiltonian for this tetraquark system is written as
 \begin{eqnarray}
 H=\sum_{k=\{Q_1, Q_2, \bar d\}}&&\left(M_k + \frac{\bm p_k^2}{2M_k} \right)-K_{c.m.} \nonumber \\
 &&+V_{Q_1 Q_2}+V_{Q_1 \bar d}+V_{Q_2 \bar d},
 \label{hamiltonian}
\end{eqnarray}
where $M_{Q_i/\bar d}$ and $\bm p_{Q_i/\bar d}$ are the mass and momentum of constituent particles, respectively. $K_{c.m.}$ is the kinetic energy of the center of mass, which is subtracted and the remained kinetic energy is for the $r_c$ and $R_c$ parts of Jacobi coordinates in Fig.~\ref{jacobi}.

For the potential between two heavy-quarks, $V_{Q_1Q_2}$, we choose the form of the ``$AP1$ potential" in Ref.~\cite{silvestre}, which reproduces the spectra of heavy mesons, color-singlet $Q \bar Q$. The $AP1$ potential contains the power-law confinement term and one-gluon-exchange term expressed as
\begin{eqnarray}
V_{Q_1 \bar Q_2}(r)&&=-\frac{\alpha}{r}+ \lambda r^{2/3} +C \nonumber \\
&&+({\bm s}_{Q_1} \cdot {\bm s}_{Q_2})  \frac{8 \kappa}{3 M_{Q_1} M_{Q_2} \sqrt{\pi}} \frac{e^{-r^2/r_0^2}}{r_0^3}, 
\label{ap1pot}
\end{eqnarray}
with the mass-dependent parameter
\begin{eqnarray}
r_0=A\left( \frac{2M_{Q_1} M_{Q_2}}{M_{Q_1} + M_{Q_2}}\right)^{-B}. 
\label{ap1sub}
\end{eqnarray}
The parameters in Eqs.~(\ref{ap1pot}) and (\ref{ap1sub}) are summarized in Table.~\ref{tableap1} with the values of effective quark masses. In addition, the calculated masses of singly and doubly heavy mesons are also summarized for the verification of this potential model.
\begin{table}[htbp]
  \centering
  \caption{Parameters and quark masses of the $AP1$ potential $V_{Q_1 \bar Q_2}(r)$ and the calculated masses of ground-state heavy mesons compared with the experimental values in Ref.~\cite{PDG}.}
  \begin{tabular}{ l  r | l  c  r } \hline\hline
\multicolumn{2}{ c |}{} 
&\multicolumn{3}{ c }{Masses ({$\rm MeV$})} \\ \cline{3-5}
\multicolumn{2}{ c |}{Parameters ($AP1$)} 
&\multicolumn{1}{ c }{Mesons} 
&\multicolumn{1}{ c }{$AP1$ \cite{silvestre}} 
&\multicolumn{1}{ c }{Exp. \cite{PDG}} \\ \hline
$\alpha$				&$0.4242$	&$D(0^-)$		&1881&1868.04\\
$\lambda$ (GeV$^{5/3}$)	&$0.3898$	&$D^*(1^-)$	&2033&2009.12\\
$C$ (GeV) 		        &$-1.1313$	&$D_s(0^-)$	&1955&1968.34\\
$\kappa$ 				&$1.8025$	&$D_s^*(1^-)$	&2107&2112.20\\ 
$A$ (GeV$^{B-1}$) 		&$1.5296$	&$B(0^-)$		&5311&5279.44\\
$B$ 					&$0.3263$	&$B^*(1^-)$	&5367&5324.70\\ 
$M_n$	(GeV)		&$0.277$		&$B_s(0^-)$	&5356&5366.88\\ 
$M_s$ 	(GeV)		&$0.553$		&$B_s^*(1^-)$	&5418&5415.40\\ 
$M_c$ 	(GeV)		&$1.819$		&$\eta_c(0^-)$	&2982&2983.90\\ 
$M_b$ 	(GeV)		&$5.206$		&$J/\psi(1^-)$	&3103&3096.90\\
					&			&$B_c(0^-)$	&6269&6274.90\\
					&			&$\eta_b(0^-)$	&9401&9398.70\\
					&			&$\Upsilon(1^-)$&9461&9460.30\\ \hline\hline
  \end{tabular}  
  \label{tableap1}
  \end{table}

Here we consider the effect of color structure expressed as $(\bm \lambda_i \cdot \bm \lambda_j)$ by the Gell-Mann matrices. As in Table.~\ref{casimir}, the value of this factor for the quark-antiquark picture inside a meson ($\bm 3 \otimes \bar {\bm 3} = \bm 1$) is two times larger than that for the two quarks with color $\bar{\bm 3}$ representation ($\bm 3 \otimes \bm 3 = \bar{\bm 3}$). 
For this reason, we assume that the $AP1$ potential in Eq.~(\ref{ap1pot}) can be generalized to the form proportional to this factor, $V_{q\bar q}(r) \propto (\bm \lambda_i \cdot \bm \lambda_j)$, and we write the potential between two heavy-quarks of color $\bar {\bm 3}$ as
\begin{eqnarray}
V_{Q_1 Q_2}(r)=\frac{1}{2}V_{Q_1 \bar Q_2} (r). 
\label{halveQQ}
\end{eqnarray}


\begin{table}[htbp]
\centering
\caption{Color structures and the values of $(\bm \lambda_i \cdot \bm \lambda_j)$, where $Q$ and $d$ denote the quark and the diquark, respectively.}
\begin{tabular}{c | c | c}\hline\hline
Two particles&Color structure&$(\bm \lambda_i \cdot \bm \lambda_j)$\\ \hline
$Q- \bar Q$, $Q- d$		&$\bm 3 \otimes \bar {\bm 3} = \bm 1$ & $-16/3$ \\
$Q- Q$, $Q- \bar d$		&$\bm 3 \otimes \bm 3 = \bar {\bm 3}$ & $-8/3$ \\ \hline\hline
\end{tabular}
\label{casimir}
\end{table}

Similarly, to construct the potential for between a heavy quark and a color $\bm 3$ antidiquark, $V_{Q_i\bar d}$, we apply the  potential diquark--heavy-quark model which gives the spectra of singly heavy baryons. For this type of potential model, we employ the ``Y-potential" in the chiral effective theory of diquarks \cite{Vdiquark}, which is firstly made from the quark model as in Refs.~\cite{yoshida, scalar}. Similar to the Eq.~(\ref{ap1pot}), this is expressed with the confinement term and one-gluon-exchange term as
\begin{eqnarray}
V_{Qd}(r)&&=-\frac{\alpha'}{r}+ \lambda' r +C'_Q \nonumber \\
 &&+({\bm s}_Q \cdot {\bm s}_d) \frac{\kappa'_Q}{M_Q M_d} \frac{\Lambda'^2}{r} \exp{(-\Lambda' r)},
\label{yoshidapot}
\end{eqnarray}
where the parameters are summarized in Table.~\ref{tableyoshida} with the calculated masses of singly heavy baryons. Note that we have adjusted the constant shift parameters $C_{c,b}$ from the original values in Ref.~\cite{Vdiquark}, so as to use the same heavy-quark mass values with the $AP1$ potential model summarized in Table.~\ref{tableap1}. For the other parameters, we use the values of diquark masses summarized in Table.~\ref{diquarkmass}. Although the Y-potential includes the spin-orbit term and the tensor term \cite{Vdiquark}, we here neglect them for simplicity.

Now we consider the color structure and the factor $(\bm \lambda_i \cdot \bm \lambda_j)$ again. From the value of $(\bm \lambda_i \cdot \bm \lambda_j)$ summarized in Table.~\ref{casimir} and the assumption that the Y-potential is proportional to the factor $(\bm \lambda_i \cdot \bm \lambda_j)$, $V_{Qd}(r) \propto (\bm \lambda_i \cdot \bm \lambda_j)$, we can write the potential between the heavy quark and the color $\bm 3$ antidiquark inside the color-singlet $T_{QQ}$ tetraquark as
\begin{eqnarray}
V_{Q_i \bar d}(r)=\frac{1}{2}V_{Qd} (r). 
\label{halveQd}
\end{eqnarray}

\begin{table}[htbp]
  \centering
  \caption{Parameters of the Y-potential and the calculated masses of ground-state singly heavy baryons compared with the experimental values in Ref.~\cite{PDG}. For the parameters, $\mu$ in the parameter $\alpha$ is the reduced mass of two interacting particles, and $C_{c,b}$ are tuned from Ref.~\cite{Vdiquark}. For the quark and diquark masses, we use the value summarized in Tables.~\ref{diquarkmass} and \ref{tableap1}.}
  \begin{tabular}{ l  r | l  c  r } \hline\hline
\multicolumn{2}{ c |}{} 
&\multicolumn{3}{ c }{Masses ({$\rm MeV$})} \\ \cline{3-5}
\multicolumn{2}{ c |}{Parameters (Y-pot.)} 
&\multicolumn{1}{ c }{Baryons} 
&\multicolumn{1}{ c }{Y-pot. \cite{Vdiquark}} 
&\multicolumn{1}{ c }{Exp. \cite{PDG}} \\ \hline
$\alpha'$				&$60/\mu$	&$\Lambda_c$ ($1/2^+$)	&2284 	&2286.46\\
$\lambda'$ (GeV$^2$)	&$0.165$		&$\Sigma_c$ ($1/2^+$)	&2451	&2453.54\\
$C'_c$ (GeV) 		        &$-0.900$		&$\Sigma_c$ ($3/2^+$)	&2513	&2518.13\\
$C'_b$ (GeV) 		        &$-0.913$		&$\Xi_c$ ($1/2^+$)		&2467	&2469.42\\
$\kappa'_c$ 			&$0.8586$	&$\Xi'_c$ ($1/2^+$)		&2581	&2578.80\\ 
$\kappa'_b$ 			&$0.6635$	&$\Xi'_c$ ($3/2^+$)		&2639	&2645.88\\ 
$\Lambda'$ (GeV) 		&$0.691$		&$\Omega_c$ ($1/2^+$)	&2699	&2695.20\\ 
					&			&$\Omega_c$ ($3/2^+$)	&2752	&2765.90\\ \cline{3-5}
					&			&$\Lambda_b$ ($1/2^+$)	&5619	&5619.60\\ 
					&			&$\Sigma_b$ ($1/2^+$)	&5810	&5813.10\\ 	
					&			&$\Sigma_b$ ($3/2^+$)	&5829	&5832.53\\ 
					&			&$\Xi_b$ ($1/2^+$)		&5796	&5794.45\\ 
					&			&$\Xi'_b$ ($1/2^+$)		&5934	&5935.02\\ 
					&			&$\Xi'_b$ ($3/2^+$)		&5952	&5953.82\\ 							
					&			&$\Omega_b$ ($1/2^+$)	&6046	&6046.10\\ 							
					&			&$\Omega_b$ ($3/2^+$)	&6063	&$\ldots$\\ \hline\hline
  \end{tabular}  
  \label{tableyoshida}
  \end{table}

To solve the  Schr$\ddot{\rm o}$dinger equation for the Hamiltonian in Eq.~(\ref{hamiltonian}), we use the Gaussian expansion method \cite{GEM1, GEM2}. Here the variational wave function of $T_{QQ}$ tetraquarks with the total spin $(J, M)$ is expressed as
\begin{eqnarray}
\Psi_{JM}&&=\sum_{c} \sum_{\beta} C_{c,\beta}
\left[ [\phi_{\nu l}^{(c)}({\bm r}_c) \psi_{NL}^{(c)}({\bm R}_c) ]_\Lambda \right. \nonumber \\
&& \left. \otimes \left[ [\chi_{1/2}(Q_1) \chi_{1/2}(Q_2)]_s \chi_{s_{\bar d}(=0, 1)}(\bar{d}) \right]_\sigma \right]_{JM},
\label{wavefunction}
\end{eqnarray}
where $\chi_{1/2}$ and $\chi_{s_{\bar d}}$ stand for the spin wave function of the heavy quark and the antidiquark, respectively. $\phi$ and $\psi$ are the Gaussian basis functions, which show the spatial wave function for ${\bm r}_c$ and ${\bm R}_c$ parts of Jacobi coordinates with the coefficients $C_{c, \beta}$. The index $\beta$ denotes the quantum numbers related to this calculation method, $\beta=\{\nu, N, l, L, \Lambda, s,\sigma\}$, where $\nu$ and $N$ are the number of basis functions, $l, L$ and $\Lambda$ are the orbital angular momentum, $ s$ and $\sigma$ are the spin of two heavy quarks and all constituent particles, respectively. Using the Gaussian expansion method, the coefficients $C_{c, \beta}$ are determined by the Rayleigh-Ritz variational principle.

In calculating the spectrum of $T_{QQ}$ tetraquarks with the antidiquark cluster, we only consider $S$-wave and $P$-wave for the orbital angular momentum. For the $S$-wave states, we calculate the masses of states which the principal quantum number $\mathcal{N}=1$ or $2$. For the $P$-wave excited states, we classify these states into three types based on the channel $c=3$ of the Jacobi coordinate in Fig.~\ref{jacobi}: the $\rho$-mode which is the excitation between two heavy quarks, the $\lambda$-mode which is that between the pair of two heavy quarks and an antidiquark, and the $\xi_d$-mode which is that between two antiquarks inside the pseudoscalar or vector antidiquark. In particular, we calculate the states summarized in Tables. \ref{statename1} and \ref{statename2}, which are represented as the spectra introduced in the next Sec.~\ref{Sec_3}. 
Note that, for the $T_{bb}$ and $T_{cc}$ tetraquarks, the quamtum numbers are restricted by the Pauli principle to $s=0$ for the $\rho$-mode states and $s=1$ for the $S$-wave and $\lambda$-mode states.

\begin{table}[htbp]
\centering
\caption{Quantum numbers of the ground and excited states of the $T_{QQ}$ tetraquarks including a scalar or an axial-vector antidiquark. 
The asterisk of the rightmost column indicates the forbidden state of $T_{bb}$ and $T_{cc}$ tetraquarks.
}
\begin{tabular}{c  | c | c  c | c | c  c | c  c}\hline\hline
State  & $\mathcal{N}$ & $l$ & $L$ & Antidiquark~($J^P$) & $s$ & $\sigma$ & $J^P$ & \\ \hline
$1S$ 			& 1 & 0 & 0 & Scalar $(0^+)$ 		&0&0&	$0^+$ &*\\
$1S$ 			& 1 & 0 & 0 & Scalar $(0^+)$ 		&1&1&	$1^+$&\\
$2S$ 			 & 2 & 0 & 0 & Scalar $(0^+)$ 		&0&0&	$0^+$&*\\ 
$2S$ 			 & 2 & 0 & 0 & Scalar $(0^+)$ 		&1&1&	$1^+$&\\ 
$\rho$ 			 & 1 & 1 & 0 & Scalar $(0^+)$ 		&0&0&	$1^-$&\\
$\rho$ 			 & 1 & 1 & 0 & Scalar $(0^+)$ 		&1&1&	$0, 1, 2^-$&*\\
$\lambda$ 		 & 1 & 0 & 1 & Scalar $(0^+)$ 		&0&0&	$1^-$&*\\
$\lambda$ 		 & 1 & 0 & 1 & Scalar $(0^+)$ 		&1&1&	$0, 1, 2^-$&\\ \hline
$1S$ 			& 1 & 0 & 0 & Axial-vector $(1^+)$ 	&0&1&	$1^+$&*\\
$1S$ 			& 1 & 0 & 0 & Axial-vector $(1^+)$ 	&1&0&	$0^+$&\\
$1S$ 			& 1 & 0 & 0 & Axial-vector $(1^+)$ 	&1&1&	$1^+$&\\
$1S$ 			& 1 & 0 & 0 & Axial-vector $(1^+)$ 	&1&2&	$2^+$&\\
$2S$ 			 & 2 & 0 & 0 & Axial-vector $(1^+)$ 	&0&1&	$1^+$&*\\ 
$2S$ 			 & 2 & 0 & 0 & Axial-vector $(1^+)$ 	&1&0&	$0^+$&\\ 
$2S$ 			 & 2 & 0 & 0 & Axial-vector $(1^+)$ 	&1&1&	$1^+$&\\ 
$2S$ 			 & 2 & 0 & 0 & Axial-vector $(1^+)$ 	&1&2&	$2^+$&\\ 
$\rho$ 			 & 1 & 1 & 0 & Axial-vector $(1^+)$ 	&0&1&	$0, 1, 2^-$&\\
$\rho$ 			 & 1 & 1 & 0 & Axial-vector $(1^+)$ 	&1&0&	$1^-$&*\\
$\rho$ 			 & 1 & 1 & 0 & Axial-vector $(1^+)$ 	&1&1&	$0, 1, 2^-$&*\\
$\rho$ 			 & 1 & 1 & 0 & Axial-vector $(1^+)$ 	&1&2&	$1, 2, 3^-$&*\\
$\lambda$ 		 & 1 & 0 & 1 & Axial-vector $(1^+)$ 	&0&1&	$0, 1, 2^-$&*\\
$\lambda$ 		 & 1 & 0 & 1 & Axial-vector $(1^+)$ 	&1&0&	$1^-$&\\
$\lambda$ 		 & 1 & 0 & 1 & Axial-vector $(1^+)$ 	&1&1&	$0, 1, 2^-$&\\
$\lambda$ 		 & 1 & 0 & 1 & Axial-vector $(1^+)$ 	&1&2&	$1, 2, 3^-$&\\ \hline \hline
\end{tabular}
\label{statename1}
\end{table}
\begin{table}[htbp]
\centering
\caption{Quantum numbers of the excited states of the $T_{QQ}$ tetraquarks including a pseudoscalar or a vector antidiquark.
The asterisk of the rightmost column indicates the forbidden state of $T_{bb}$ and $T_{cc}$ tetraquarks.
}
\begin{tabular}{c  | c | c  c | c | c  c | c  c }\hline\hline
State  & $\mathcal{N}$ & $l$ & $L$ & Antidiquark~($J^P$) & $s$ & $\sigma$ & $J^P$ & \\ \hline
$\xi_P$ 			& 1 & 0 & 0 & Pseudoscalar $(0^-)$ 	&0&0&	$0^-$&*\\
$\xi_P$ 			& 1 & 0 & 0 & Pseudoscalar $(0^-)$	&1&1&	$1^-$&\\
$\xi_P\rho$ 		& 1 & 1 & 0 & Pseudoscalar $(0^-)$ 	&0&0&	$1^+$&\\
$\xi_P\rho$ 		& 1 & 1 & 0 & Pseudoscalar $(0^-)$ 	&1&1&	$0, 1, 2^+$&*\\
$\xi_P\lambda$ 	 & 1 & 0 & 1 & Pseudoscalar $(0^-)$ 	&0&0&	$1^+$&*\\
$\xi_P\lambda$ 	 & 1 & 0 & 1 & Pseudoscalar $(0^-)$ 	&1&1&	$0, 1, 2^+$&\\ \hline
$\xi_V$ 			& 1 & 0 & 0 & Vector $(1^-)$ 		&0&1&	$1^-$&*\\
$\xi_V$ 			& 1 & 0 & 0 & Vector $(1^-)$ 		&1&0&	$0^-$&\\
$\xi_V$ 			& 1 & 0 & 0 & Vector $(1^-)$ 		&1&1&	$1^-$&\\
$\xi_V$ 			& 1 & 0 & 0 & Vector $(1^-)$ 		&1&2&	$2^-$&\\
$\xi_V\rho$ 		 & 1 & 1 & 0 & Vector $(1^-)$ 		&0&1&	$0, 1, 2^+$&\\
$\xi_V\rho$ 		 & 1 & 1 & 0 & Vector $(1^-)$ 		&1&0&	$1^+$&*\\
$\xi_V\rho$ 		 & 1 & 1 & 0 & Vector $(1^-)$ 		&1&1&	$0, 1, 2^+$&*\\
$\xi_V\rho$ 		 & 1 & 1 & 0 & Vector $(1^-)$ 		&1&2&	$1, 2, 3^+$&*\\
$\xi_V\lambda$ 	 & 1 & 0 & 1 & Vector $(1^-)$ 		&0&1&	$0, 1, 2^+$&*\\
$\xi_V\lambda$ 	 & 1 & 0 & 1 & Vector $(1^-)$ 		&1&0&	$1^+$&\\
$\xi_V\lambda$ 	 & 1 & 0 & 1 & Vector $(1^-)$ 		&1&1&	$0, 1, 2^+$&\\
$\xi_V\lambda$ 	 & 1 & 0 & 1 & Vector $(1^-)$ 		&1&2&	$1, 2, 3^+$&\\ \hline \hline
\end{tabular}
\label{statename2}
\end{table}
  
\section{Results} \label{Sec_3}

\subsection{Spectrum of $T_{bb}$ and $T_{cc}$ tetraquarks}

\begin{figure*}[htb]
  \includegraphics[clip,width=1.77\columnwidth]{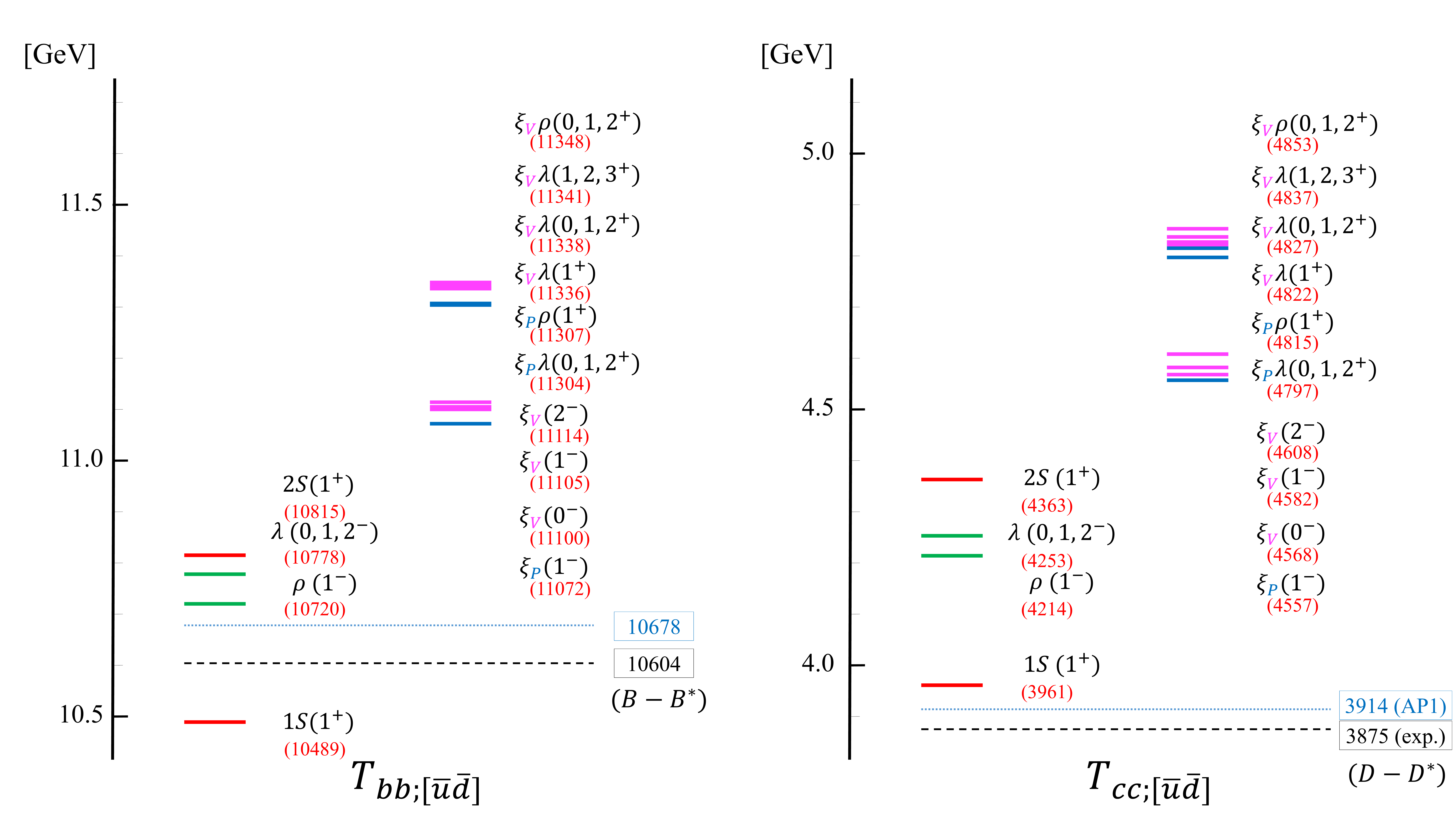}
  \caption{The energy spectra of nonstrange $T_{bb}$ and $T_{cc}$ tetraquarks with the flavor $\bm 3$ antidiquarks. The colors of lines show the types of states and constituent antidiquarks, such as red for the $S$-wave state with the scalar antidiquark, green for the $P$-wave state with the scalar antidiquark, blue for the ones with the pseudoscalar antidiquark, and magenta for the ones with the vector antidiquark. The dashed and dotted lines are the thresholds relevant to the strong decay of the ground state, where the black-dashed ones are calculated from the experimental values of the meson masses in Ref.~\cite{PDG} and the blue-dotted ones are from the meson masses calculated in the $AP1$ potential model, Eqs.~(\ref{ap1pot}) and (\ref{ap1sub}).} 
  \label{figureTbbccud}
  \includegraphics[clip,width=1.77\columnwidth]{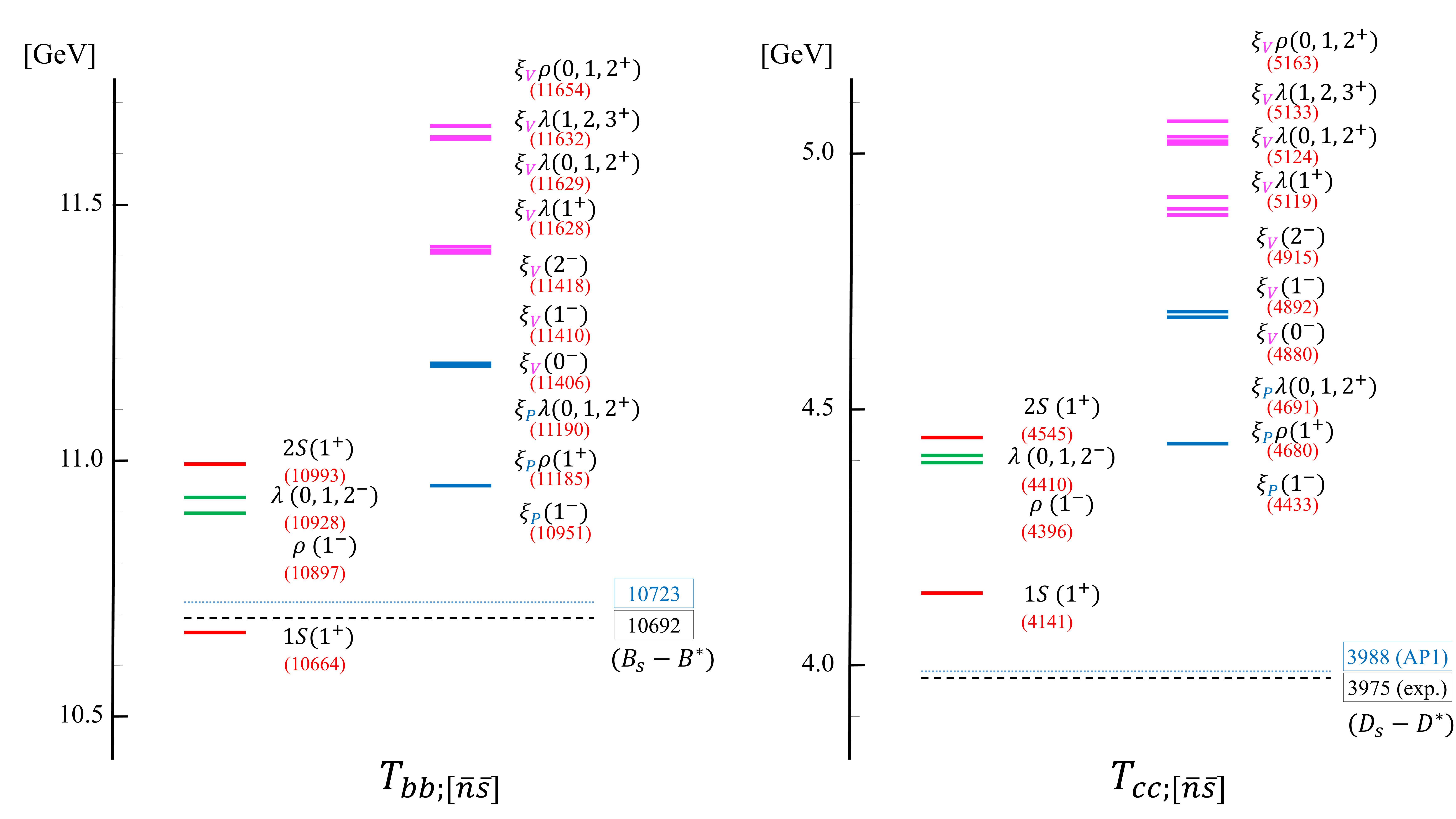}
  \caption{The energy spectra of strange $T_{bb}$ and $T_{cc}$ tetraquarks with the flavor $\bm 3$ antidiquarks. The notations are the same as Fig.~\ref{figureTbbccud}.} 
  \label{figureTbbccns}
\end{figure*}
\begin{figure*}[htb]
  \includegraphics[clip,width=1.85\columnwidth]{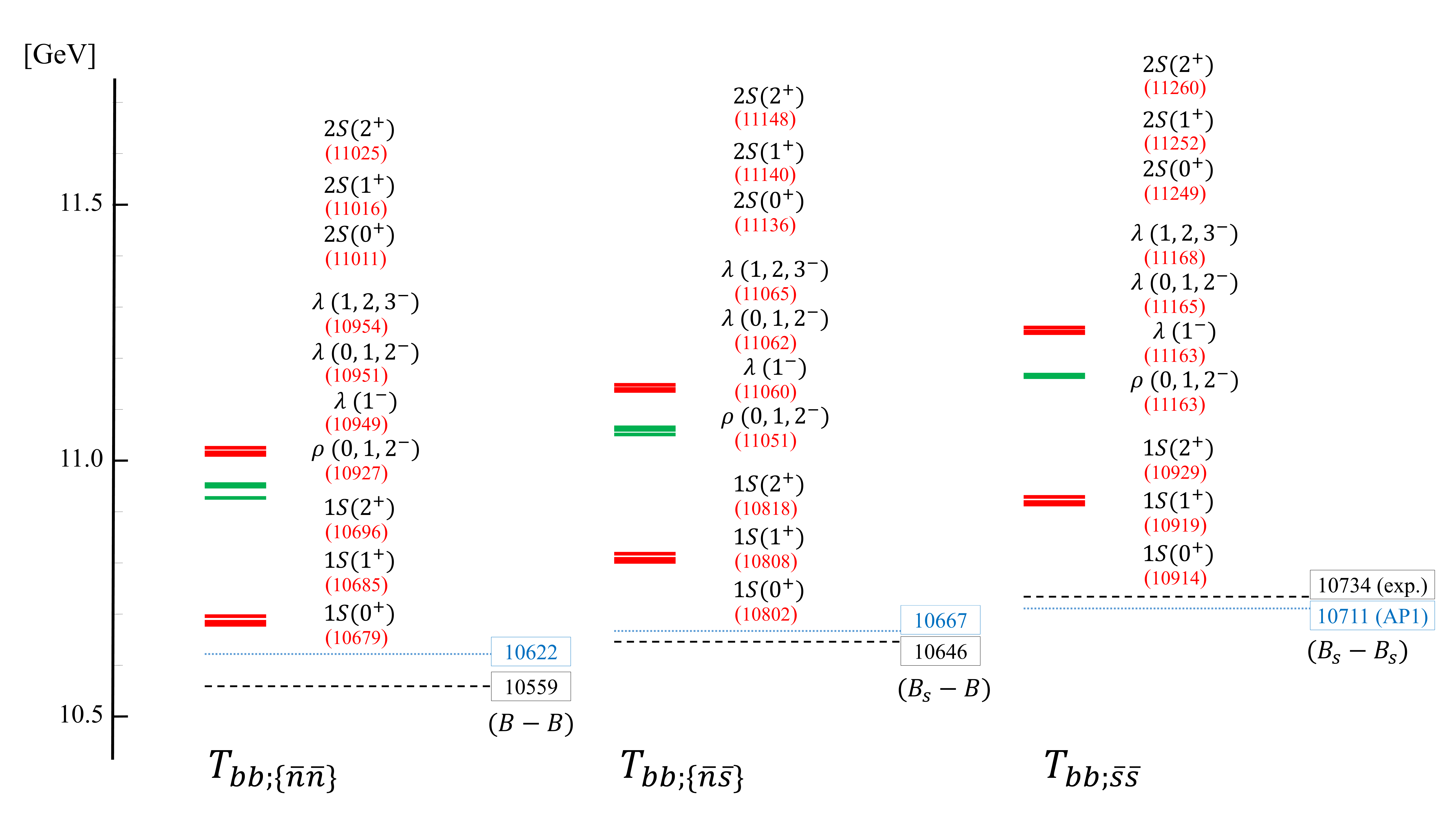}
  \caption{The energy spectra of $T_{bb}$ tetraquarks with the flavor $\bar {\bm 6}$ antidiquark. The red and green states are the $S$-wave and $P$-wave states with the axial-vector antidiquark, respectively. The other notations are the same as Fig.~\ref{figureTbbccud}.} 
  \label{figureTbbaxial}
  \includegraphics[clip,width=1.85\columnwidth]{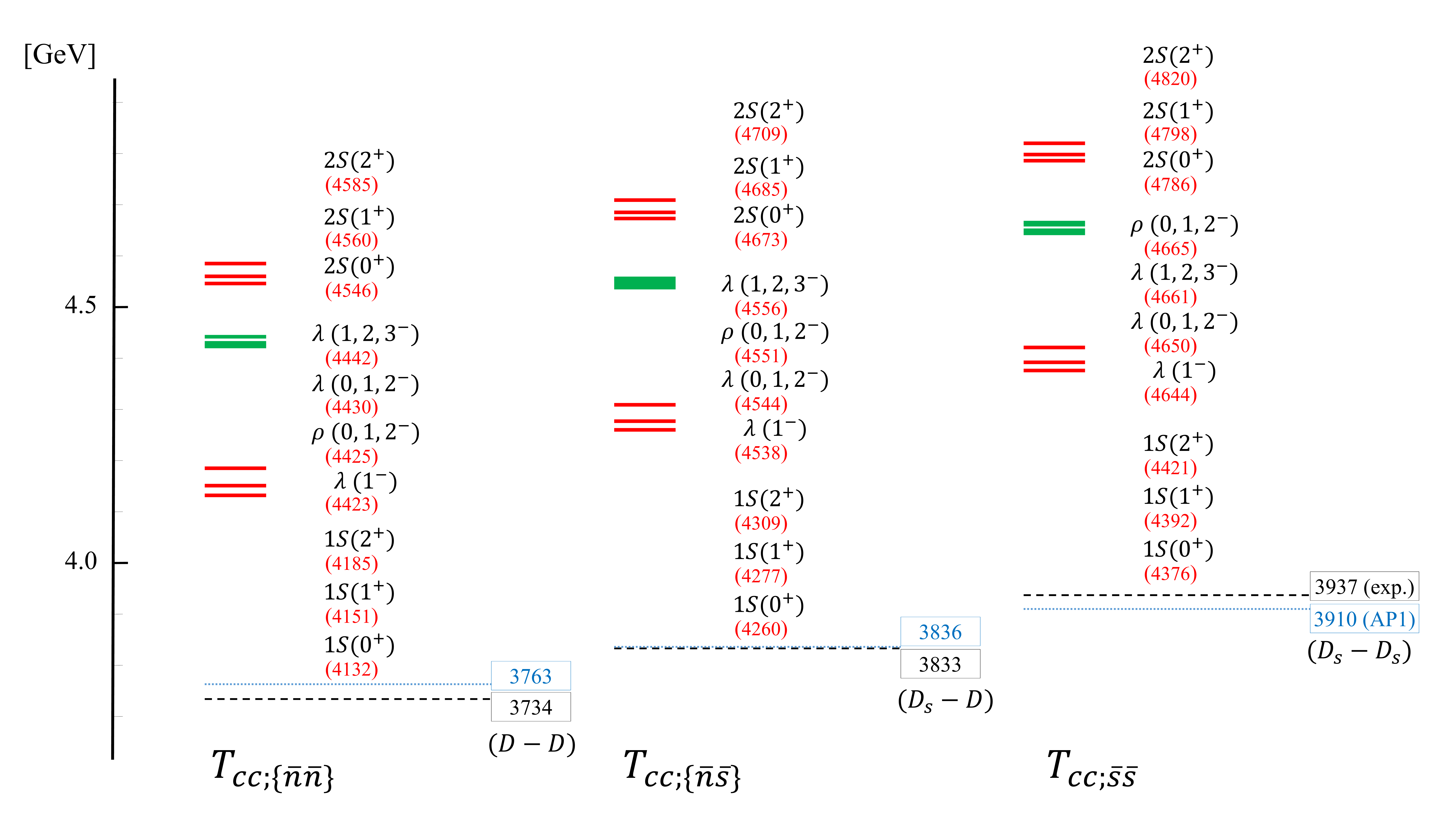}
  \caption{The energy spectra of $T_{cc}$ tetraquarks with the flavor $\bar {\bm 6}$ antidiquark. The other notations are the same as Fig.~\ref{figureTbbaxial}.} 
  \label{figureTccaxial}
\end{figure*}

Here we discuss the spectra of the $T_{bb}$ and $T_{cc}$ tetraquarks, in which two constituent heavy quarks are the same. 
The energy spectra of these tetraquarks are shown in Figs.~\ref{figureTbbccud}, \ref{figureTbbccns}, \ref{figureTbbaxial}, and \ref{figureTccaxial}. Each figure is classified by the flavor of constituent antidiquarks and the heavy quarks.

Figs. \ref{figureTbbccud} and \ref{figureTbbccns} are the spectra of the strangeness $S=0$ and $+1$ tetraquarks, respectively. Here the flavor of all the constituent antidiquarks is $\bm 3$, meaning the scalar, pseudoscalar, and vector antidiquarks. In particular, the red and green lines stand for the masses of tetraquarks including the scalar antidiquark, which correspond to the $S$-wave states and the $P$-wave states, respectively. For the other colors, the blue and magenta lines, indicate the tetraquarks including the pseudoscalar antidiquark and the vector antidiquark, respectively.

About the spectra in Fig.~\ref{figureTbbccud}, all the tetraquarks including the scalar antidiquarks are lighter than those containing the pseudoscalar or vector antidiquarks. On the other hand, in Fig.~\ref{figureTbbccns}, not all the tetraquarks including the scalar antidiquarks are lighter than the others. In detail, when a constituent antidiquark has an strange antiquark,, the masses of $\xi_P$-mode states become lighter than those of $2S$ states and approach to the lightest $P$-wave states ($\rho$- and $\lambda$-mode states). This is caused by the inverse mass hierarchy of pseudoscalar diquarks in Eq.~(\ref{inverse}). Moreover, all the tetraquarks containing the $\xi_V$-mode state become heavier than the other states by the inclusion of a strange antiquark, and this is caused by the enhanced mass hierarchy of vector diquarks in Eq.~(\ref{vec-sqmassdiff}).

The thresholds for the strong decay of the $1^+$ ground states is given by the sum of $J^P=0^-$ and $1^-$ singly heavy mesons ($B$ or $D$). Note that the $1^+$ ground state is not allowed to decay into two $0^-$ mesons.
In Figs.~\ref{figureTbbccud} and \ref{figureTbbccns}, we show two sets of threshold values, one (exp.) given by the experimental meson masses in the Particle Data Group \cite{PDG} (black-dashed lines), and another (AP1) from the calculation with the potential $AP1$ in Eqs.~(\ref{ap1pot}) and (\ref{ap1sub}) (blue-dotted lines). One sees that the $1S$ ground states of $T_{bb}$ tetraquarks are below the $B_{(s)}-B^*$ thresholds, so that they are bound states and do not decay by the strong interaction. From the experimental thresholds, their binding energies are $115$ MeV for the $T_{bb;\bar u \bar d}$ and $28$ MeV for the $T_{bb;\bar n \bar s}$. On the other hand, no bound state appears in the spectra of $T_{cc}$ tetraquarks.


Next we discuss the spectra of the tetraquarks containing the axial-vector antidiquarks. These are illustrated in Figs.~\ref{figureTbbaxial} and \ref{figureTccaxial}, which show the spectrum of $T_{bb}$ and $T_{cc}$ tetraquarks, respectively. The red and green lines stand for the $S$-wave states and the $P$-wave states, respectively. In these figures, the number of strange antiquark is increasing from left to right.

Comparing three spectra in each figure, the masses of these tetraquarks in each state become heavier as the number of the strange antiquark increases, and the mass difference between $T_{QQ;\bar n\bar n}$ and $T_{QQ;\bar n\bar s}$ is almost equal to that between $T_{QQ;\bar n\bar s}$ and $T_{QQ;\bar s\bar s}$. This behavior comes from the generalized Gell-Mann--Okubo mass formula for the axial-vector diquarks in Eq.~(\ref{axialgomass}).

The thresholds shown in Figs.~\ref{figureTbbaxial} and \ref{figureTccaxial} are for the $0^+$ tetraquarks, and are the total mass of two singly heavy mesons with $J^P=0^-$. Although the $1S$ ground state with $J^P=0^+$ are the lightest, they are above the thresholds and therefore unstable. Discrepancy between the tetraquark mass and the threshold is even larger for the strange tetraquarks.


About the spin-spin potential terms in the two potential models, Eqs.~(\ref{ap1pot}) and (\ref{yoshidapot}), both of them are in inverse proportion to the masses of two interacting particles. Therefore, in the heavy-quark limit ($M_{Q_i} \rightarrow \infty$), these terms disappear, and the splitted states by the spin-spin interaction become degenerate. This is called the heavy-quark spin (HQS) multiplet which appears frequently in the tetraquark spectra. For example, in Figs.~\ref{figureTbbaxial} and \ref{figureTccaxial}, the $1S$ and $2S$ states of tetraquarks with $J^P=0^+, 1^+$, and $2^+$ have the HQS triplet structure. Comparing these structures in Figs.~\ref{figureTbbaxial} and \ref{figureTccaxial}, we can see that the mass differences for $ T_{bb}$ tetraquarks are smaller than those for $T_{cc}$ tetraquarks, which is caused by the bottom quark being heavier than the charm quark.

\subsection{Spectrum of $T_{cb}$ tetraquarks}

\begin{figure*}[htb]
  \includegraphics[clip,width=1.85\columnwidth]{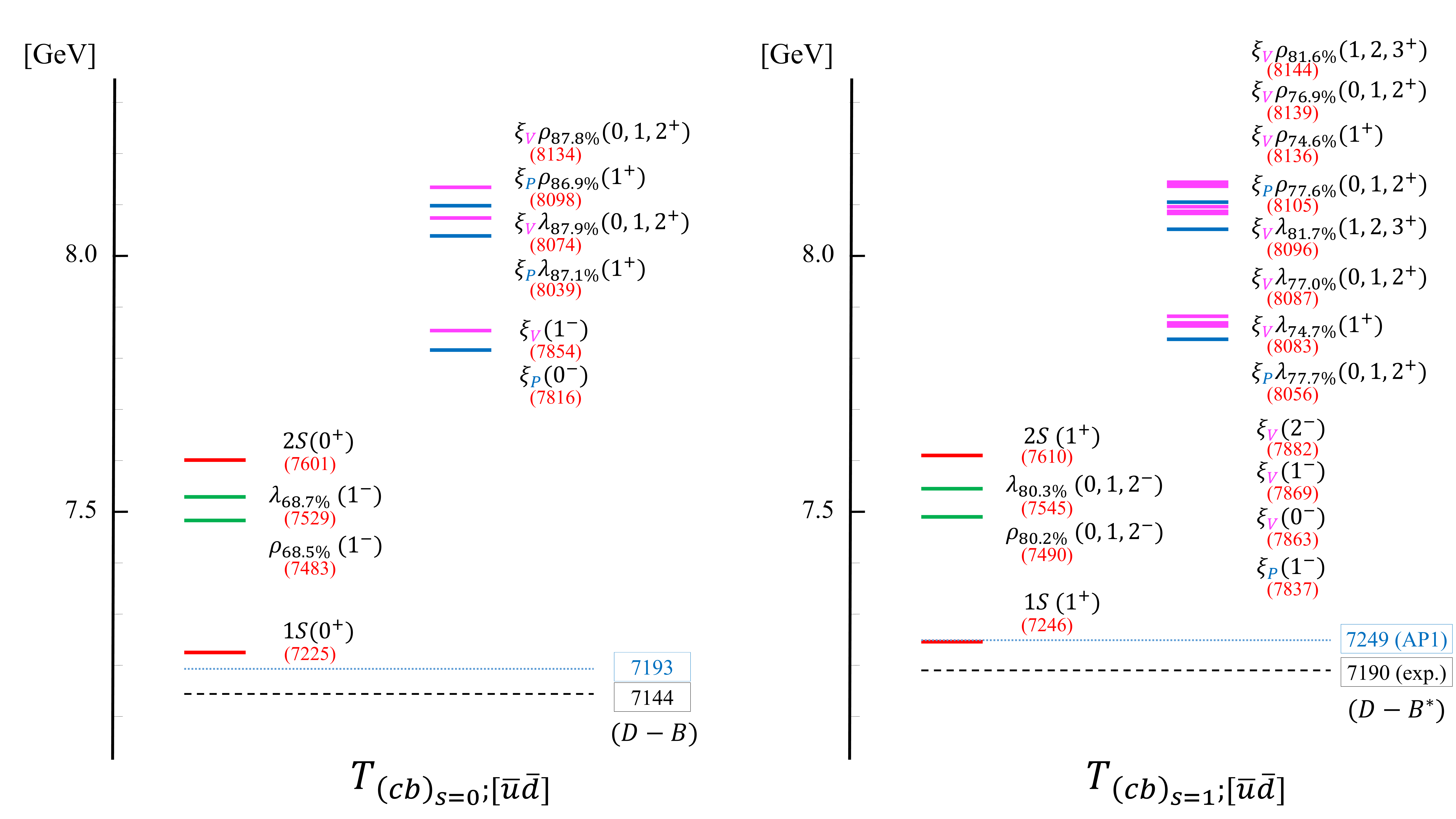}
  \caption{The energy spectra of nonstrange $T_{cb}$ tetraquarks with the flavor $\bm 3$ antidiquarks, classified by the total spin of heavy quarks, $s=0$, or $1$. The other notations are the same as Fig.~\ref{figureTbbccud}.} 
  \label{figureTcbud}
  \includegraphics[clip,width=1.85\columnwidth]{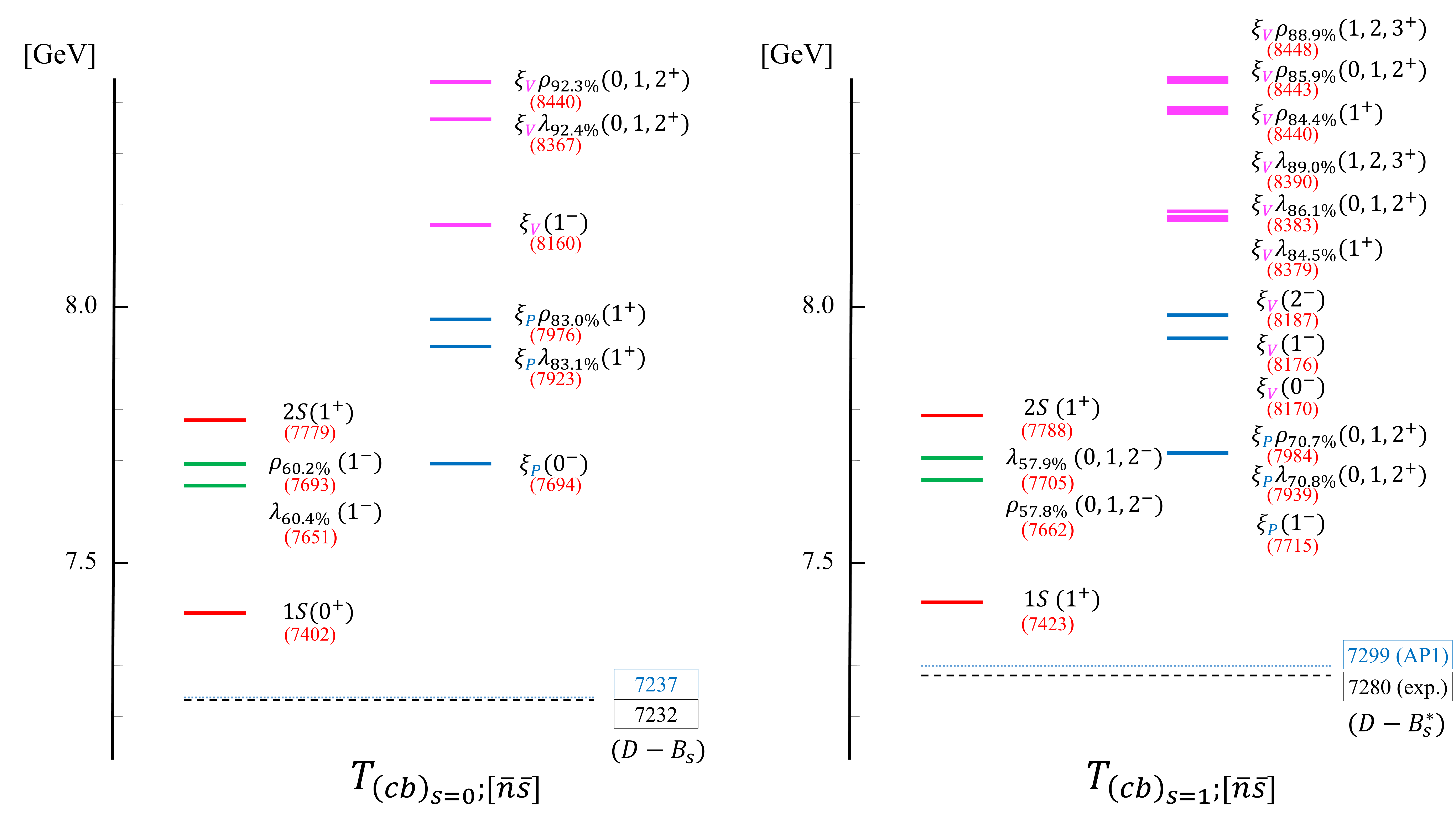}
  \caption{The energy spectra of strange $T_{cb}$ tetraquarks with the flavor $\bm 3$ antidiquarks, classified by the total spin of heavy quarks, $s=0$, or $1$. The other notations are the same as Fig.~\ref{figureTbbccud}.} 
  \label{figureTcbns}
\end{figure*}
\begin{figure*}[htb]
  \includegraphics[clip,width=1.85\columnwidth]{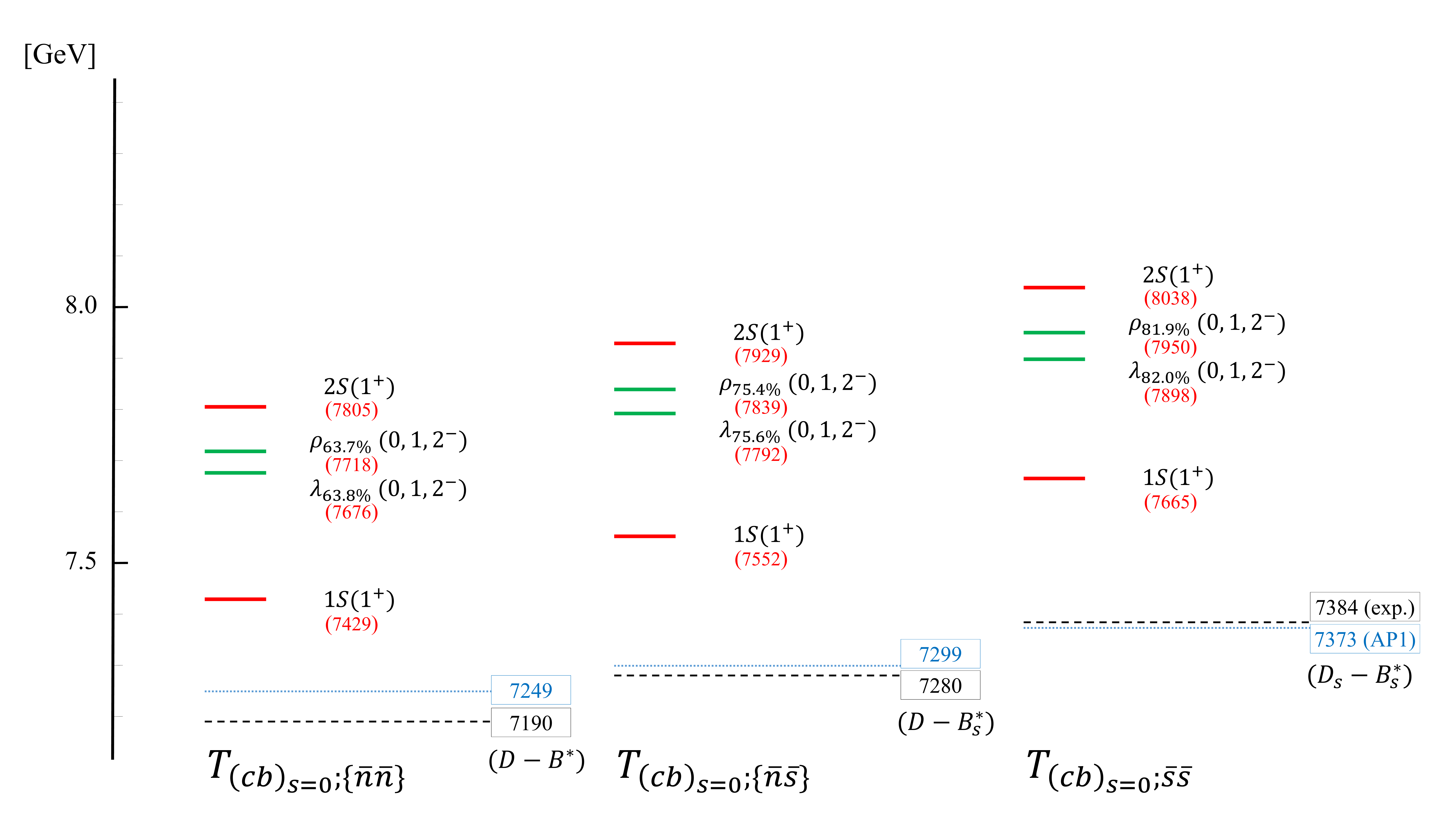}
  \caption{The energy spectra of $T_{cb}$ tetraquarks with the flavor $\bar {\bm 6}$ antidiquarks, for the total spin of heavy quarks, $s=0$. The other notations are the same as Fig.~\ref{figureTbbaxial}.} 
  \label{figureTcb0axial}
  \includegraphics[clip,width=1.85\columnwidth]{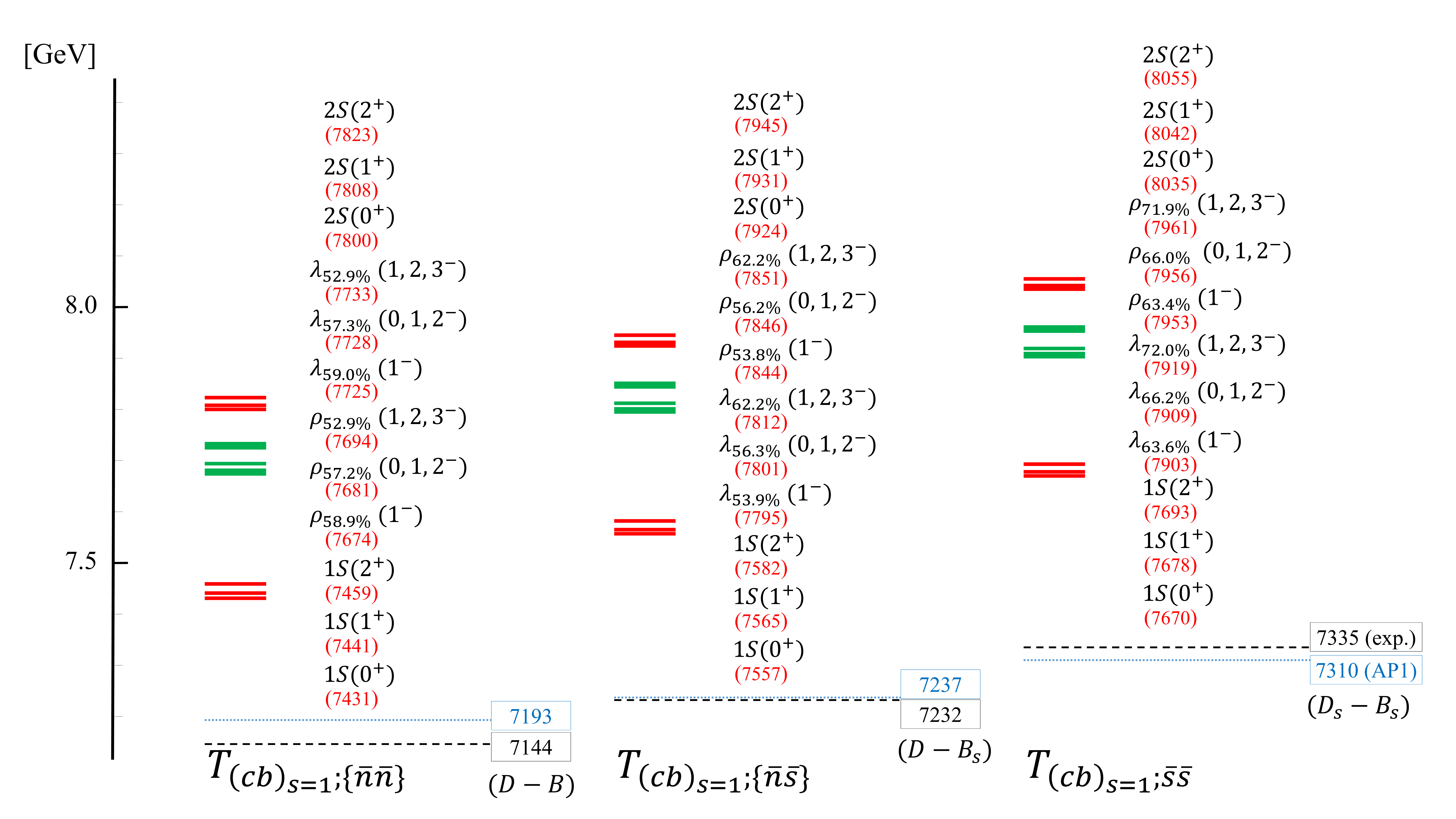}
  \caption{The energy spectra of $T_{cb}$ tetraquarks with the flavor $\bar {\bm 6}$ antidiquarks, for the total spin of heavy quarks, $s=1$. The other notations are the same as Fig.~\ref{figureTbbaxial}.} 
    \label{figureTcb1axial}
\end{figure*}

We here discuss the spectra of $T_{cb}$ tetraquarks. Since two constituent heavy quarks are different from each other, the spin of two heavy quarks, $s$, can take two kinds of values as $s=0$ and $1$. The energy spectra of these tetraquarks are shown in Figs.~\ref{figureTcbud}, \ref{figureTcbns}, \ref{figureTcb0axial}, and \ref{figureTcb1axial}, classified by the flavor of constituent antidiquarks and the quantum number $s$. 
Figs.~\ref{figureTcbud} and \ref{figureTcbns} are the spectra of the strangeness $S=0$ and $+1$ tetraquarks with a flavor $\bm 3$ antidiquark, respectively. The four colors of lines stand for the types of states and constituent antidiquarks, which is the same as in Figs.~\ref{figureTbbccud} and \ref{figureTbbccns}.
Also, Figs.~\ref{figureTcb0axial} and \ref{figureTcb1axial} are the spectra of the $T_{cb}$ tetraquarks with $s=0$ and $1$, including a flavor $\bar{\bm 6}$ axial-vector antidiquark, respectively. The two colors of lines stand for the $S$-wave states and the $P$-wave states, which is the same as Figs.~\ref{figureTbbaxial} and \ref{figureTccaxial}.

The spectra of $T_{cb}$ tetraquarks are similar to those of $T_{bb}$ and $T_{cc}$ tetraquarks when the constituent antidiquarks are the same. Although there are less states for the $T_{cb}$ tetraquarks with $s=0$, we can see the effect of chiral effective theory of diquarks: the inverse mass hierarchy of pseudoscalar diquarks and the enhanced mass hierarchy of vector diquarks in Figs.~\ref{figureTcbud} and \ref{figureTcbns}, and the generalized Gell-Mann--Okubo mass formula for the axial-vector diquarks in Figs.~\ref{figureTcb0axial} and \ref{figureTcb1axial}.

In the states given in Tables. \ref{statename1} and \ref{statename2}, one sees that some $P$-wave states have the same spin quantum numbers, such as $\rho$ and $\lambda$ modes with $s=\sigma=0$ or 1.
These states may be mixed by the spin-spin interaction in $T_{cb}$. (Note that the mixing is not allowed in $T_{bb}$ and $T_{cc}$ due to the Pauli principle.)
The subscripts of $\rho$ and $\lambda$ characters mean the probabilities of each state, calculated by the two-body density distribution between two heavy quarks. 
For example, about two $P$-wave states with green lines in Fig.~\ref{figureTcbud}, the $\rho$- and $\lambda$-mode states mix each other. As a result, in the lower state, the probability of the $\rho$-mode state becomes $68.5\%$, and the remained $31.5\%$ is that of the $\lambda$-mode state.

As with the $T_{bb}$ and $T_{cc}$ spectra, the black-dashed (exp.) and the blue-dotted (AP1) lines in Figs.~\ref{figureTcbud}, \ref{figureTcbns}, \ref{figureTcb0axial}, and \ref{figureTcb1axial} are the relevant two-meson thresholds. We find no bound $T_{cb}$ tetraquarks except the $T_{cb;\bar u \bar d}$ tetraquark with $s=1$ at 3 MeV below the AP1 threshold in Fig.~\ref{figureTcbud}. Therefore, we conclude that most of the $T_{cb}$ tetraquarks with the antidiquark cluster are unstable in strong decays. 


\subsection{Binding energies of lowest $T_{QQ}$ tetraquark states}

\begin{table}[htbp]
  \centering
  \caption{The binding energies of the $T_{QQ}$ tetraquark ground states including a scalar antidiquark, which are based on two types of thresholds: type exp from the experimental masses of mesons in Ref.~\cite{PDG}, and type AP1 from the theoretical masses of mesons given by the $AP1$ potential model. The character $I$ shows for the isospin.}
  \begin{tabular}{  c r  | c c } \hline\hline
    \multicolumn{1}{c }{Tetraquark}
  &\multicolumn{1}{c |}{State}
  &\multicolumn{2}{c }{Binding energy [MeV]} \\ \hline
    \multicolumn{1}{ c }{$T_{QQ;\bar{q}\bar{q}}$}
  &\multicolumn{1}{r |}{$ I(J^P)$}
  &\multicolumn{1}{c }{type exp}
  &\multicolumn{1}{c }{type AP1} \\ \hline
 $T_{bb;\bar{u}\bar{d}}$			&$ 0(1^+)$ 	&$115$	&$189$		\\ 
 $T_{bb;\bar{n}\bar{s}}$			&$ 1/2(1^+)$ 	&$28$  	&$59$	\\ 
 $T_{cc;\bar{u}\bar{d}}$			&$ 0(1^+)$ 	&$(-86)$	&$(-47)$		\\ 
 $T_{cc;\bar{n}\bar{s}}$			&$ 1/2(1^+)$ 	&$(-166)$  &$(-153)$	\\  
 $T_{cb;\bar{u}\bar{d}}$			&$ 0(0^+)$ 	&$(-81)$  &$(-32)$	\\
 $T_{cb;\bar{n}\bar{s}}$			&$ 1/2(0^+)$ 	&$(-170)$  &$(-165)$	\\
 $T_{cb;\bar{u}\bar{d}}$			&$ 0(1^+)$ 	&$(-56)$  &$3$	\\
 $T_{cb;\bar{n}\bar{s}}$			&$ 1/2(1^+)$ 	&$(-143)$  &$(-124)$	\\
  \hline \hline
  \end{tabular}  
  \label{boundTQQ}
  \end{table}

We summarize the binding energies of the lowest-lying states in Table~\ref{boundTQQ}. Two lowest states of $T_{bb}$ tetraquarks are well below the threshold and expected to be stable.

On the other hand, we have no bound $T_{cc}$ state. It may seem contradictory to the recent observation of $T_{cc}^+$, at about 300 keV below the $D^{*+}-D^0$ threshold by the LHCb collaboration \cite{LHCb:2021vvq, LHCb:2021auc}. As the observed state is very close to the threshold, it may couple strongly to the $D^{*+}-D^0$ state and thus have a molecule-like structure. In contrast, the tetraquark in our diquark picture does not dissociate into $D-D$, and therefore it does not couple to loosely bound $D^{(*)}-D$ states. In other words, our model can reproduce only tightly bound tetraquark states. Thus our results may not be inconsistent with the LHCb observation.

In Table.~\ref{tabletheorycomparison}, we also summarize the stability of lowest $T_{QQ}$ tetraquark states given by some previous works. Most of the studies in this table predicted that the non-strange and singly antistrange $T_{bb}$ tetraquarks, $T_{bb;\bar{u}\bar{d}}$ and $T_{bb;\bar{n}\bar{s}}$, are stable, which is consistent with our results. On the other hand, the other tetraquark systems seem to be unstable or uncertain. 

\begin{table*}[tb!]
\centering
\caption{Comparison between theoretical studies for the stability of doubly heavy tetraquarks.
``S'', ``US'', and ``ND'' denote ``stable'' (below the threshold), ``unstable'' (above the threshold), and ``not determined'' (in other words, not conclusive within theoretical uncertainties), respectively.
For a similar summary, see Ref.~\cite{Cheng:2020wxa}.}
\label{table: stability}

\begin{tabular}{l | ccc  ccc  ccc}\hline\hline
Reference&$T_{cc;\bar{u}\bar{d}}$&$T_{cc;\bar{n}\bar{s}}$&$T_{cc;\bar{s}\bar{s}}$&$T_{bb;\bar{u}\bar{d}}$&$T_{bb;\bar{n}\bar{s}}$&$T_{bb;\bar{s}\bar{s}}$&$T_{cb;\bar{u}\bar{d}}$&$T_{cb;\bar{n}\bar{s}}$&$T_{cb;\bar{s}\bar{s}}$\\
\hline

\hline
\multicolumn{1}{l}{Quark-level models}&\multicolumn{9}{c}{} \\ \hline

\hline
%
Lipkin (1986)~\cite{Lipkin:1986dw}
&ND&&&S&&&&&\\
ZouZou et al. (1986)~\cite{Zouzou:1986qh}
&S&&&S&&&S&&\\
Carlson-Heller-Tjon (1988)~\cite{Carlson:1987hh}
&S&&&S&&&ND&&\\ 
%
%
Semay--Silvestre-Brac (1994)~\cite{Semay:1994ht}
&ND&&&S&S&&ND&&\\ 
Pepin et al. (1997)~\cite{Pepin:1996id}
&S&&&S&&&&&\\
%
%
%
%
Janc-Rosina (2004)~\cite{Janc:2004qn}
&S&&&S&&&&&\\
%
%
%
Ebert et al. (2007)~\cite{Ebert:2007rn}
&US&US&US&S&US&US&US&US&US\\
Zhang-Zhang-Zhang (2008)~\cite{Zhang:2007mu}
&US&&US&S&&US&&&\\ 
Vijande-Valcarce-Barnea (2009)~\cite{Vijande:2009kj}
&S&&&S&&&&&\\ 
Yang et al. (2009)~\cite{Yang:2009zzp}
&ND&&&S&&&&&\\ 
Feng-Guo-Zou (2013)~\cite{Feng:2013kea}
&S&&&S&&&S&&\\
Karliner-Rosner (2017)~\cite{Karliner:2017qjm}
&US&&&S&&&ND&&\\
Eichten-Quigg (2017)~\cite{tetraeichten}
&US&US&US&S&S&US&US&US&US\\
Park-Noh-Lee (2019)~\cite{Park:2018wjk}
&US&&&S&S&&US&&\\
Deng-Chen-Ping (2020)~\cite{Deng:2018kly}
&S&US&US&S&S&US&S&US&US\\
Yang-Ping-Segovia (2020a)~\cite{Yang:2019itm}
&S&&&S&&&S&&\\
Tan-Lu-Ping (2020)~\cite{Tan:2020ldi}
&S&&&S&&&S&&\\
L\"u-Chen-Dong (2020)~\cite{Lu:2020rog}
&US&US&US&S&US&US&US&US&US\\
Braaten-He-Mohapatra (2021)~\cite{tetrabraaten}
&US&US&US&S&S&US&US&US&US\\ 
Yang-Ping-Segovia (2020b)~\cite{Yang:2020fou}
&&&US&&&US&&&US\\
Meng et al. (2021)~\cite{tetrahiyama}
&S&&&S&S&&S&&\\
(This work) 
&US&US&US&S&S&US&ND&US&US\\
\hline

\hline
\multicolumn{1}{l}{Chromomagnetic interaction models}&\multicolumn{9}{c}{} \\ \hline


\hline
%
Lee-Yasui (2009)~\cite{Lee:2009rt}
&S&S&&S&S&&S&US&\\
Luo et al. (2017)~\cite{Luo:2017eub}
&S&S&US&S&S&US&S&S&US\\
Cheng et al. (2020)~\cite{Cheng:2020nho}
&&&&&&&ND&US&\\
Cheng et al. (2021)~\cite{Cheng:2020wxa}
&US&US&US&S&S&US&ND&US&US\\
Weng-Deng-Zhu (2022)~\cite{Weng:2021hje}
&S&ND&US&S&S&ND&S&ND&ND\\
Guo et al. (2022)~\cite{Guo:2021yws}
&S&S&US&S&S&ND&S&S&US\\
\hline

\hline
\multicolumn{1}{l}{QCD sum rules}&\multicolumn{9}{c}{} \\ \hline


\hline
%
Navarra-Nielsen-Lee (2007)~\cite{Navarra:2007yw}
&ND&&&S&&&&&\\
Du et al. (2013)~\cite{Du:2012wp}
&&US&US&S&S&S&&&\\
Chen-Steele-Zhu (2014)~\cite{Chen:2013aba}
&&&&&&&ND&&S\\
Wang (2018)~\cite{Wang:2017uld}
&ND&ND&&S&S&&&&\\
Wang-Yan (2018)~\cite{Wang:2017dtg}
&ND&ND&ND&&&&&&\\
Agaev et al. (2019)~\cite{Agaev:2018khe}
&&&&S&&&&&\\
Agaev-Azizi-Sundu (2020)~\cite{Agaev:2019kkz}
&&&&&&&ND&&\\
Agaev et al. (2020, 2021)~\cite{Agaev:2019lwh, Agaev:2020zag}
&&&&&S&&&&\\
Wang-Chen (2020)~\cite{Wang:2020jgb}
&&&&&&&&S&\\
\hline

\hline
\multicolumn{1}{l}{Lattice QCD}&\multicolumn{9}{c}{} \\ \hline


\hline
%
Francis et al. (2017)~\cite{Francis:2016hui}
&&&&S&S&&&&\\
Francis et al. (2019)~\cite{Francis:2018jyb}
&&&&S&S&&S&&\\
Junnarkar-Mathur-Padmanath (2019)~\cite{tetralatticebbcc}
&S&S&US&S&S&ND&&&\\
Leskovec et al. (2019)~\cite{Leskovec:2019ioa}
&&&&S&&&&&\\
Hudspith et al. (2020)~\cite{tetralatticebc}
&&&&S&S&&US&S&\\
Mohanta-Basak (2020)~\cite{Mohanta:2020eed}
&&&&S&&&&&\\
\hline\hline
\end{tabular}
\label{tabletheorycomparison}
\end{table*}

\subsection{Density distribution of $T_{bb}$ bound states}

\begin{figure}[htbp]
\centering
  \includegraphics[width=1.05\columnwidth]{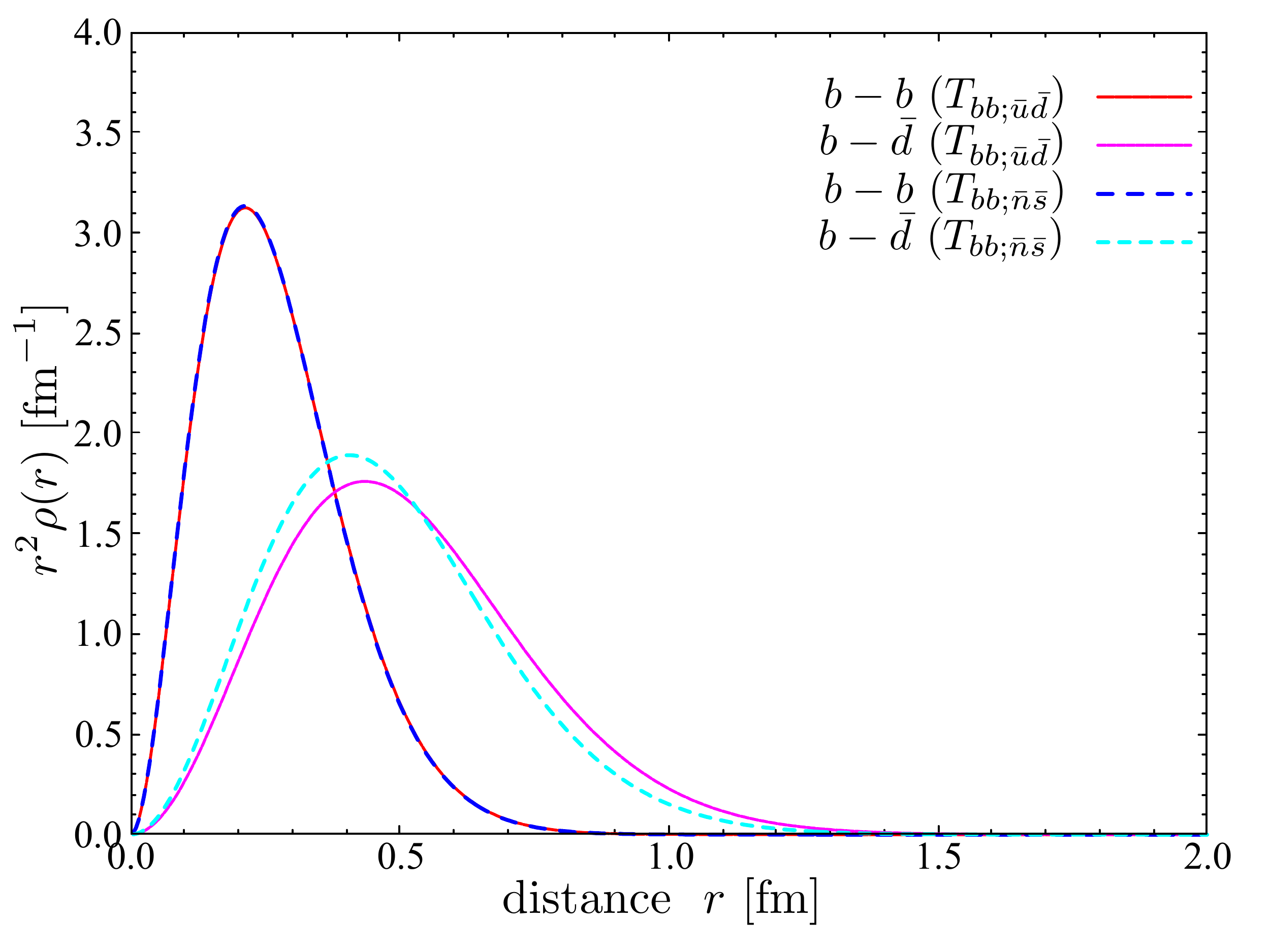}
  \caption{Density distributions between two constituent bottom quarks ($b-b$) and between a bottom quark and an antidiquark ($b-\bar d$) inside the bound states of $T_{bb}$ tetraquarks. The solid lines are for the strangeness $S=0$ tetraquark, and the dashed lines are for the strangeness $S=+1$ tetraquark, respectively.} 
  \label{figureTbbdensity}
\end{figure}

Now we discuss the density distributions between two constituent particles in each of the $T_{bb}$ tetraquark bound states. The two-body density distribution is expressed by the variational wave function in Eq.~(\ref{wavefunction}) as
\begin{eqnarray}
\rho(r_c)=\int d\hat{\bm r}_c d{\bm R}_c |\Psi_{JM}({\bm r}_c, {\bm R}_c)|^2,
\label{dens}
\end{eqnarray}
where $r_c=|{\bm r}_c|$ and $\hat{\bm r}_c$ are the distance and the angular parts of the relative coordinate for two particles in channel $c$, respectively. From this equation (\ref{dens}), we can give the density distribution between two bottom quarks with channel $c=3$, and that between a bottom quark and a scalar antidiquark with channel $c=1, 2$. The functions $r^2\rho(r)$ of each pair are shown in Fig.~\ref{figureTbbdensity} for both $T_{bb}$ tetraquarks with the strangeness $S=0$ and $+1$.

We see that the distance between the pair of two bottom quarks ($b-b$) is nearer than the pair of a bottom quark and an antidiquark ($b-\bar d$) for both strangeness $S=0$ and $+1$. This is due to the suppression of kinetic energy which is inversely proportional to the reduced mass of two interacting particles. 
Comparing with the different strangeness $S$, the red solid line and the blue dashed line are almost same so that there is no difference between the pair $b-b$. On the other hand, for between the pair $b-\bar d$, the nonstrange antidiquark seems to be more extended from a bottom quark than that including one strange antiquark, because the scalar diquark becomes heavier by including a strange quark.

\begin{table}[htbp]
  \centering
  \caption{The rms distances of the $T_{bb}$ tetraquark bound states.}
  \begin{tabular}{ c  r | c  c } \hline\hline
    \multicolumn{1}{l }{Tetraquark}
  &\multicolumn{1}{c |}{State}
  &\multicolumn{2}{c }{rms distance $\sqrt{\hat r^2}$ [fm] } \\ \hline
    \multicolumn{1}{c }{$T_{bb;{\bar q}{\bar q}}$}
  &\multicolumn{1}{r |}{$ I(J^P)$}
  &\multicolumn{1}{c }{$~~{b-b}~~$  }
  &\multicolumn{1}{c }{${b-\bar d}$ } \\ \hline 
 $T_{bb;{\bar u}{\bar d}}$	&$ 0(1^+)$ 	&0.30&0.56\\ 
 $T_{bb;{\bar n}{\bar s}}$&$ 1/2(1^+)$ 	&0.30&0.53\\ \hline \hline
  \end{tabular}  
  \label{tablerms}
  \end{table}

Also we summarize the root-mean-square (rms) distances of between two particles in Table.~\ref{tablerms}, which is given by the density distribution function $\rho(r)$ as
\begin{eqnarray}
\sqrt{\hat r_c^2}=\sqrt{\left( \int r_c^2 \rho(r_c) r_c^2 dr_c \right) / \left( \int \rho(r_c) r_c^2 dr_c\right)} .
\label{rms}
\end{eqnarray}
For the rms distance between a bottom quark and an antidiquark ($b-\bar d$), these values are similar to that between a bottom quark and a diquark inside the ground state of $\Lambda_b$ ($\simeq 0.55$ fm) and $\Xi_b$ ($\simeq 0.51$ fm)~\cite{scalar}.
Although the interaction between a heavy quark and an antidiquark is half as large as that between a heavy quark and a diquark from Eq.~(\ref{halveQd}), its effect is mostly suppressed by the interaction between two bottom quarks.
 
\subsection{$T_{QQ}$ tetraquarks toward chiral restoration}


In the chiral effective theory based on the linear sigma model, the spontaneous chiral symmetry breaking is controlled by the vacuum expectation value of $\langle \Sigma \rangle$. It is interesting to see how the tetraquark masses are modified when the value of $\langle \Sigma \rangle$ changes. Such a situation may be realized in nature when the tetraquarks are placed or produced in the hot/dense matter, where chiral symmetry tends to be restored. According to Ref.~\cite{Vdiquark}, when we multiply the chiral symmetry breaking factor $x$ to $\langle \Sigma \rangle$, where its range is $0 \le x \le 1$, we can express the masses of diquarks from the chiral symmetry restored phase ($x=0$) to the ordinary vacuum state ($x=1$).

For example, the masses of nonstrange diquarks are given by
\begin{eqnarray}
&& M_{ud}(0^+) = \sqrt{m^2_{S0} - (x+\epsilon) m^2_{S1} - x^2 m^2_{S2}},\label{Schiral}\\
&& M_{ud}(0^-) = \sqrt{m^2_{S0} + (x+\epsilon) m^2_{S1} + x^2 m^2_{S2}},\label{Pchiral}\\
&& M_{nn}(1^+) = \sqrt{m^2_{V0} + x^2( m^2_{V1} + 2m^2_{V2})}, \label{Achiral}\\
&& M_{ud}(1^-) = \sqrt{m^2_{V0} + x^2(-m^2_{V1} + 2m^2_{V2})}, \label{Vchiral}
\end{eqnarray}
in which only the mass of scalar diquark $M_{ud}(0^+)$ becomes heavier and the others become lighter with approaching to the chiral symmetry restored phase $(x \rightarrow 0)$. 

\begin{figure}[htbp]
\centering
  \includegraphics[width=1.05\columnwidth]{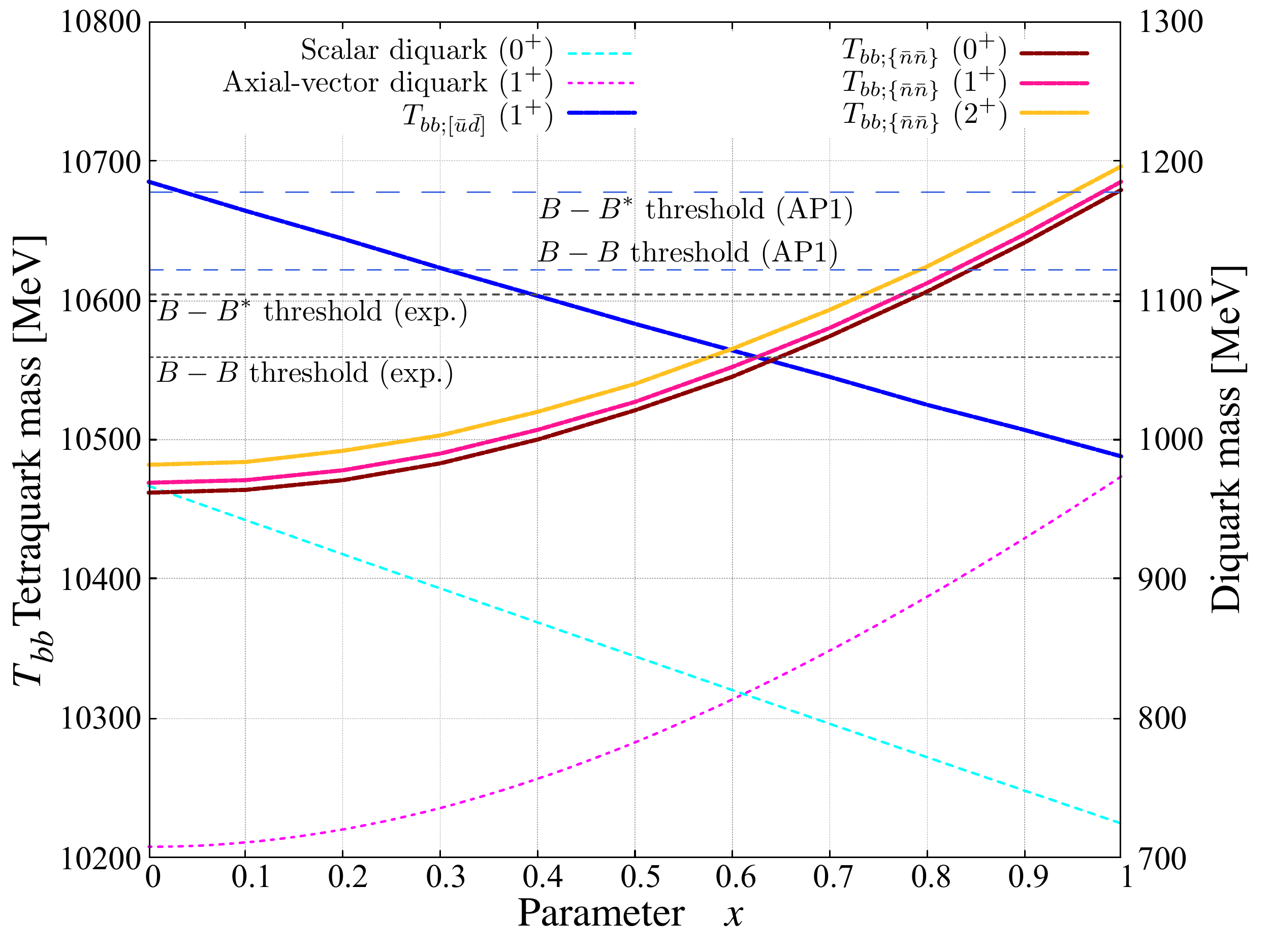}
  \caption{Dependence of the masses of nonstrange $T_{bb}$ tetraquarks (left scale) on the chiral symmetry breaking strength $x$. Dependence of the masses of the diquarks (right scale) are also shown for reference. 
  }
  \label{Tbbchiral}
\end{figure}
\begin{figure}[htbp]
\centering
  \includegraphics[width=1.05\columnwidth]{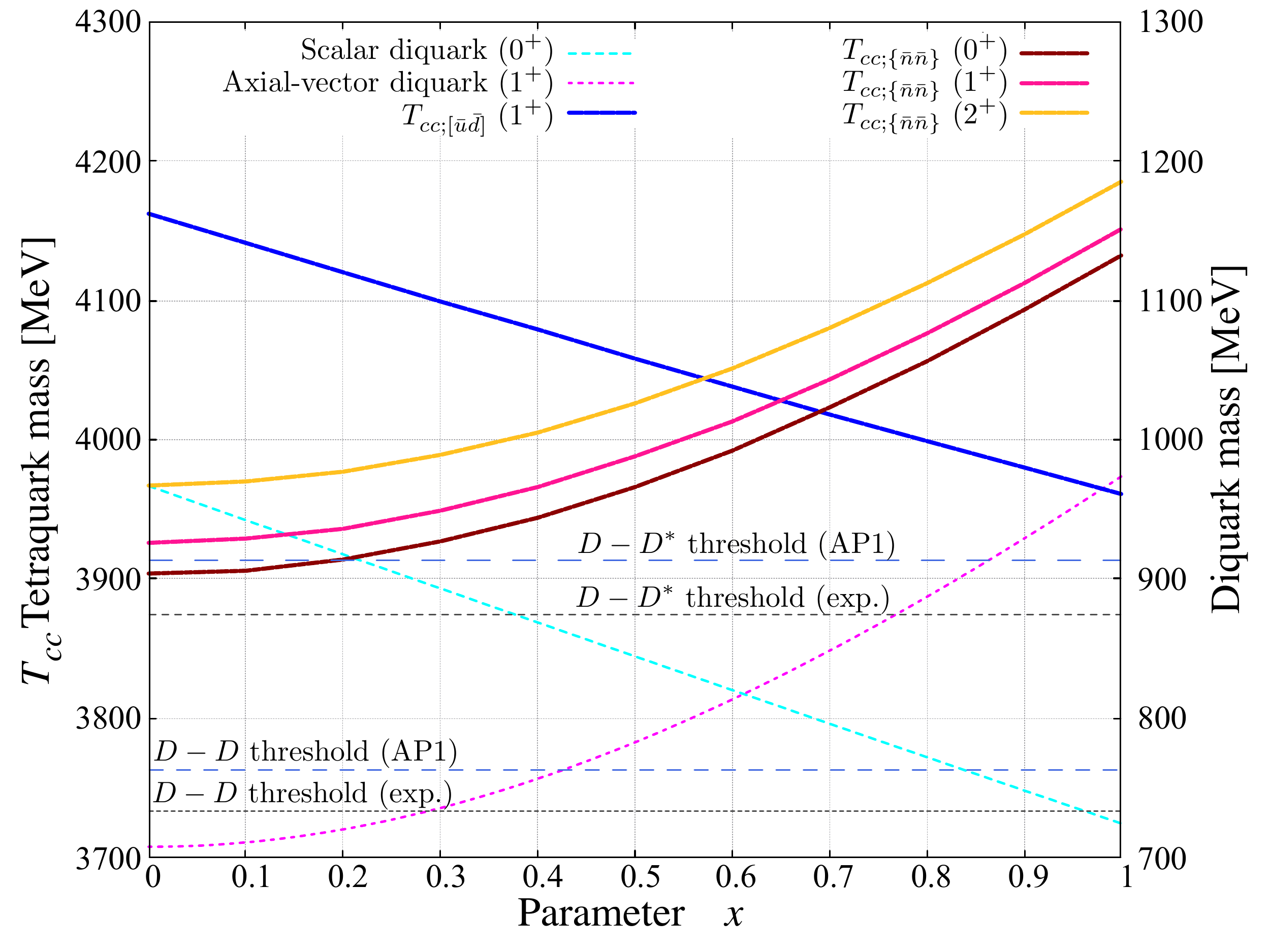}
  \caption{Dependence of the masses of nonstrange $T_{cc}$ tetraquarks on the chiral symmetry breaking strength $x$. 
  }
  \label{Tccchiral}
\end{figure}
\begin{figure}[htbp]
\centering
  \includegraphics[width=1.05\columnwidth]{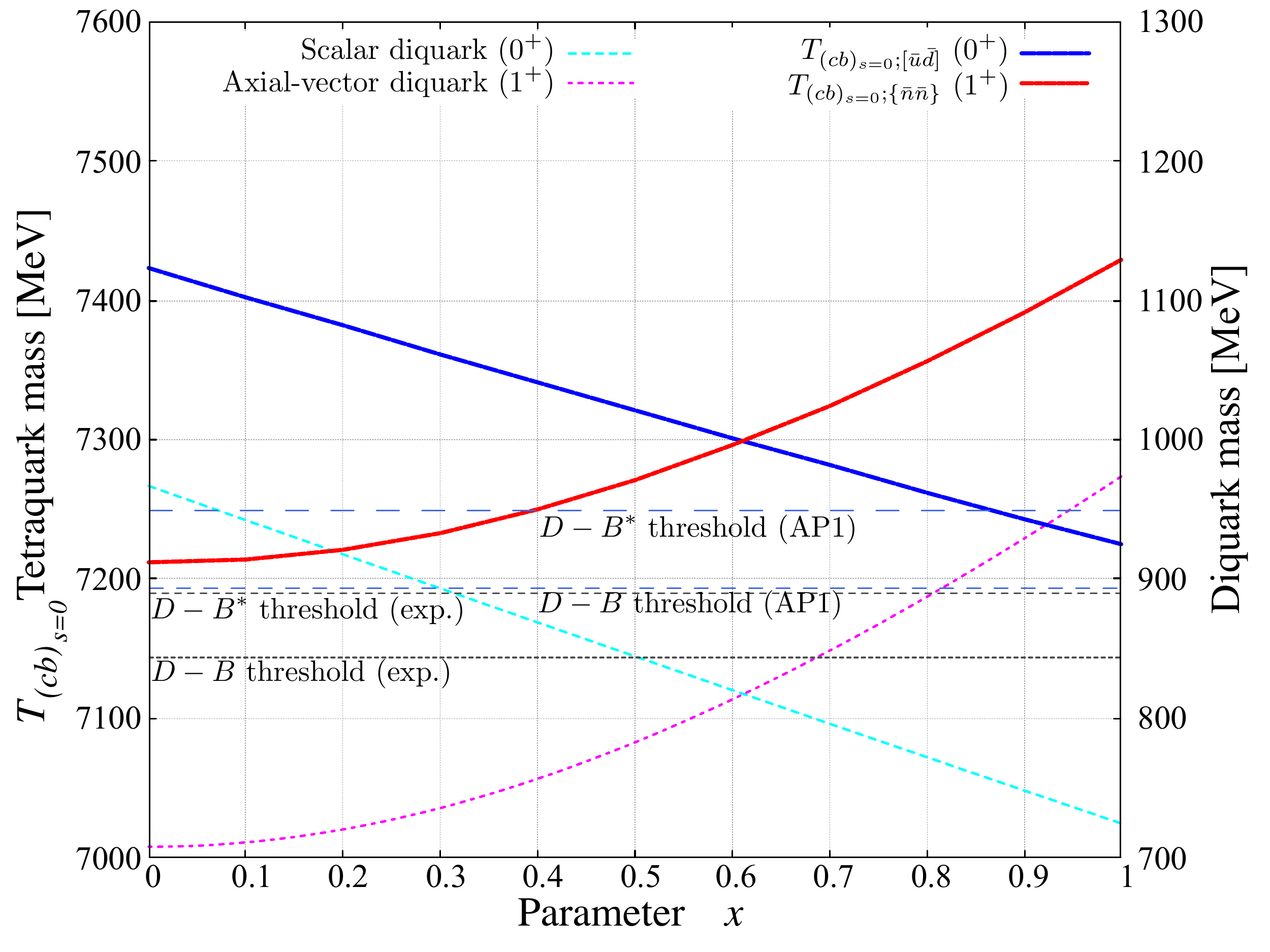}
  \caption{Dependence of the masses of nonstrange $T_{cb}$ tetraquarks with $s=0$ on the chiral symmetry breaking strength $x$. 
  }
  \label{Tcb0chiral}
\end{figure}
\begin{figure}[htbp]
\centering
  \includegraphics[width=1.05\columnwidth]{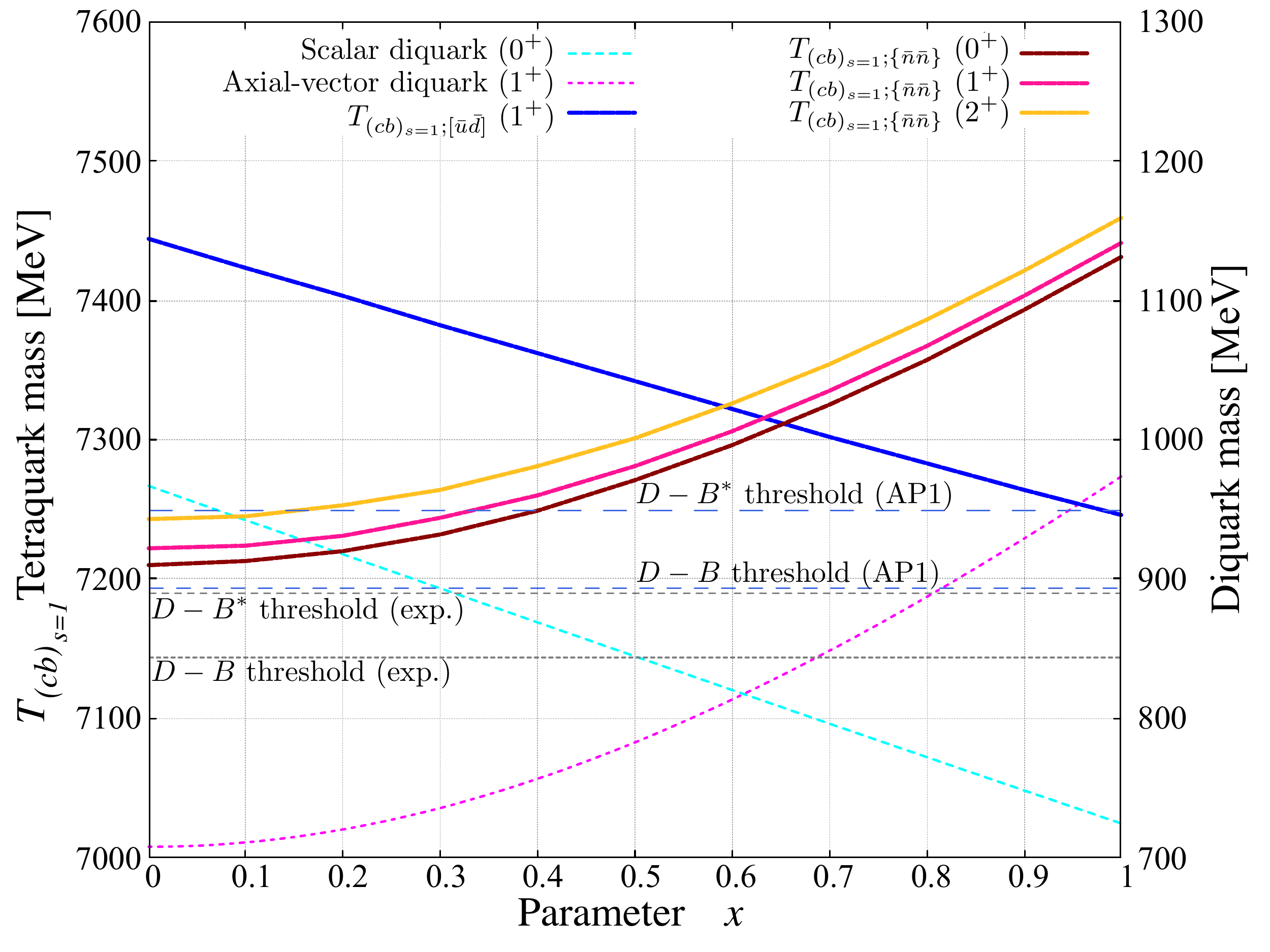}
  \caption{Dependence of the masses of nonstrange $T_{cb}$ tetraquarks with $s=1$ on the chiral symmetry breaking strength $x$. 
  }
  \label{Tcb1chiral}
\end{figure}

Figs.~\ref{Tbbchiral}, \ref{Tccchiral}, \ref{Tcb0chiral}, and \ref{Tcb1chiral} show the change of masses of $T_{QQ}$ tetraquarks by the factor $x$, which are all $1S$ ground states including a scalar or an axial-vector antidiquark. For all tetraquarks in these figures, these masses change, which is similar to the mass of an antidiquark they include. As the masses of scalar and axial-vector diquarks cross at around $x=0.6$ and are reversed, the crossing of masses of $T_{QQ}$ tetraquarks also occurs near $x=0.6$. The tetraquarks including an axial-vector antidiquark become lighter than those including a scalar antidiquark. 
In other words, the isospin of stable $T_{QQ}$ tetraquarks changes from $0$ to $1$ as the chiral symmetry is restored.
In particular, if we assume that meson-meson thresholds do not change by the degree of chiral symmetry breaking, $T_{bb}$ tetraquarks with a constituent axial-vector antidiquark are below the thresholds when the parameter $x$ is less than $0.6$. 

In the chiral symmetry restored phase, the masses of nonstrange vector diquark and axial-vector diquark are degenerate with the mass value $m_{V0}$ according to Eqs.~(\ref{Achiral}) and (\ref{Vchiral}). Therefore, in the chiral symmetry restored phase, not only the positive-parity states but also the negative-parity states with the $\xi_V$-mode are below the thresholds for the nonstrange $T_{bb}$ tetraquark. On the other hand, the masses of nonstrange pseudoscalar diquark and scalar diquark are not degenerate due to the $U_A(1)$ anomaly, and the pseudoscalar diquark is about 300 MeV heavier than the scalar diquark even at $x=0$ \cite{Vdiquark}. Thus there are no $\xi_P$-mode state below the thresholds for the nonstrange $T_{bb}$ tetraquark.

\section{Summary} \label{Sec_4}
In this paper, we have studied the spectrum of doubly heavy $T_{QQ}$ tetraquarks in the potential chiral-diquark model. Applying the chiral effective theory of diquarks and the potential quark-diquark model \cite{CETdiquark, scalar, Vdiquark}, 
we have calculated the lower energy spectrum and the wave functions of $S$-wave and $P$-wave $T_{QQ}$ tetraquarks. 
Here we have employed the Gaussian expansion method \cite{GEM1, GEM2} for solving the three-body system consisting of two heavy quarks and one antidiquark. We have also investigated the behavior of the ground-state energies of the nonstrange tetraquarks under the chiral restoration.

For the $T_{QQ}$ tetraquarks including an antidiquark, we have obtained the following:
\begin{enumerate}
\item[(i)] For the ground states of the $T_{QQ}$ tetraquarks, we have found that the $T_{bb}$ tetraquarks including a scalar antidiquark are stable in strong decays with the binding energies, 115 MeV for $T_{bb;\bar u \bar d}$ and 28 MeV for $T_{bb; \bar n \bar s}$. No other states, such as $T_{cc}$ and $T_{cb}$ tetraquarks, are significantly below the two-meson thresholds and thus not stable.
\item[(ii)] For the excited states of the $T_{QQ}$ tetraquarks, we have found the inverse mass hierarchy in states with the $\xi_P$-mode and the enhanced mass hierarchy in states with the $\xi_V$-mode. As a result, for the singly antistrange tetraquarks, there are one more $P$-wave states between $1S$ and $2S$ states compared with the nonstrange tetraquarks.
\item[(iii)] For two stable bound states of $T_{bb}$ tetraquarks, we have investigated the structure by calculating the density distributions and rms distances from the wave functions. We found that the size of nonstrange and singly antistrange $T_{bb}$ tetraquarks are similar to each other and have less difference by the existence of an strange antiquark. 
\item[(iv)] For the ground states of the nonstrange $T_{QQ}$ tetraquarks, we have investigated the change of masses by the degree of chiral symmetry breaking. Here we used the mass formulas of non-strange diquarks expressed with the chiral symmetry breaking factor $x$, Eqs.~(\ref{Schiral}) -- (\ref{Vchiral}). We found that the axial-vector antidiquark makes the $T_{QQ}$ tetraquarks more stable than the scalar antidiquark near the chiral symmetry restored phase.
\end{enumerate}


Finally, we comment on possible improvements of our model.
In this work, we have treated the antidiquark as a point-like particle, where the distance between two light antiquarks inside the antidiquark is neglected. It will be important to take into account the finite size of diquarks and improve the heavy-quark--diquark potential model, as done by Refs.~\cite{Kumakawa:2017ffl, Kumakawa:2021ujz}. In such a model, the parameters of the improved heavy-quark--diquark potential model will be fixed to reproduce the low-lying masses of the singly heavy baryons, and the results of the excited $T_{QQ}$ tetraquark states would be improved. 

The chiral effective field theory constructed in Refs.~\cite{CETdiquark, scalar, Vdiquark}, includes only the scalar, axial-vector, pseudoscalar, and vector diquarks. As summarized in Table.~\ref{diopchiral}, all the diquarks considered in this work have color $\bar{\bm 3}$, so that the energy spectra of $T_{QQ}$ tetraquarks including such a diquark are related only to the color ${\bm 3}-\bar{\bm 3}$ interaction between the light antidiquark and the heavy diquark. 
On the other hand, another type is the color $\bm 6$ diquark. If we consider the color $\bm 6$ diquarks in the chiral effective theory of diquarks, the energy spectra of $T_{QQ}$ tetraquarks is related also to the color $\bar{\bm 6}-{\bm 6}$ interaction.
According to Refs.~\cite{Deng:2018kly, Lu:2020rog, Chen:2021tnn, Hyodo:2012pm, Luo:2017eub, Hyodo:2017hue, Cheng:2020nho, Cheng:2020wxa, Weng:2021hje, Guo:2021yws}, the energy spectra of $T_{QQ}$ tetraquarks including the color $\bar{\bm 6}$ antidiquark would appear above the ground states given by the color $\bm 3$ scalar antidiquarks. Also, tensor diquarks with color $\bar{\bm 3}$ and flavor $\bm 6$, or with color $\bm 6$ and flavor $\bar{\bm 3}$, would be the new source for the excited states of $T_{QQ}$ tetraquarks. Such studies are left for future works.


\section*{Acknowledgments}
We would like to thank Yan-Rui Liu for helpful discussions.
This work was supported by Grants-in-Aid for Scientific Research No. JP20K03959 (M.O.), No. JP21H00132 (M. O.), No. JP17K14277 (K. S.), and No. JP20K14476 (K. S.), and for JSPS Fellows No. JP21J20048 (Y. K.) from Japan Society for the Promotion of Science.

\clearpage


\bibliography{ref-tetraquark}
\end{document}